\documentclass[]{aastex631}
\usepackage {subfigure}
\usepackage{appendix}
\usepackage{rotating}
\usepackage{longtable}
\usepackage {multirow}
\usepackage{pifont}

\shortauthors{Deng \& Li}
\graphicspath{{./}{figures/}}

\begin{document}

\title{Are there Black Hole Symbiotic X-ray Binaries?}

\author[0000-0002-1398-2694]{Zhu-Ling Deng}
\affil{School of Astronomy and Space Science, Nanjing University, Nanjing 210023, China; lixd@nju.edu.cn }
\affil{Key Laboratory of Modern Astronomy and Astrophysics (Nanjing University), Ministry of Education, Nanjing 210023, China}
\author[0000-0002-0584-8145]{Xiang-Dong Li}
\affil{School of Astronomy and Space Science, Nanjing University, Nanjing 210023, China; lixd@nju.edu.cn }
\affil{Key Laboratory of Modern Astronomy and Astrophysics (Nanjing University), Ministry of Education, Nanjing 210023, China}

\begin{abstract}
While there are over a dozen known neutron star (NS) symbiotic X-ray binaries (SyXBs) in the Galaxy, none SyXBs containing a black hole (BH) have been detected. We address this problem by incorporating binary population synthesis and the accretion properties of BHs fed by the wind from red giant companions. We investigate the impact of different supernova mechanisms, kick velocity distributions and wind velocities on the formation of both NS and BH SyXBs. Our simulations show that the number of BH SyXBs is at most one-sixth that of NS SyXBs in the Galaxy provided that the common envelope efficiency parameter $\alpha\sim 0.3-5$. And less than $\sim 10$ of BH SyXBs could be detectable in X-ray, considering their low radiation efficiencies. These findings indicate a scarcity of BH SyXBs in the Galaxy.
    
\end{abstract}


\section{Introduction}

Symbiotic X-ray binaries (SyXBs) are a subgroup of binary star systems consisting of an accreting neutron star (NS) or black hole (BH) with a giant companion, typically of late type (K1-M8) \citep[][for reviews]{Luna2013,Kuranov2015,Sazonov2020,Bahramian2023}. SyXBs exhibit relatively low X-ray luminosities, ranging from $\sim 10^{32}$ to $10^{36}$ erg\,s$^{-1}$. Since the first SyXB discovered over four decades ago \citep{Davidsen1977}, about a dozen SyXBs have been detected to date, all of which harbor NS accretors \citep[see][Table 1]{Yungelson2019}. The measured orbital periods of known SyXBs span from 151 days \citep[CGCS 5926,][]{Masetti2011} to 1161 days \citep[GX 1+4,][]{Hinkle2006,IMM2017}. In two sources, the orbital eccentricities have been measured, that is, $e=0.1$ for GX 1+4 \citep{Hinkle2006} and $e=0.26\pm 0.15$ for 4U 1700+24 \citep{Galloway2002}. 

Theoretically, \citet{Lv2012} developed a binary population synthesis (BPS) model for SyXBs in the Galaxy, taking account of both stellar/binary evolution and the interaction between the magnetized NS and the accreting wind material. Their simulations, considering varying wind velocities and the angular momentum distribution in the shell of the matter settling on to a NS, predicted a SyXB population of approximately $100-1000$ systems currently residing in our galaxy. \citet{Yungelson2019} updated the \citet{Lv2012} model, incorporating advancements in the settling accretion model \citep{Shakura2018} and updated SyXB observations. Their revised estimates suggested that there are slightly less than 40 SyXBs per $10^{10}M_{\odot}$ (or $\sim 240$ SyXBs in the Galaxy).

To date, no BH SyXBs have been identified \citep{Bahramian2023}. Considering the fact that there are at least hundreds of NS SyXBs predicted in the Milky Way, this raises the interesting question: are there or how many are there BH SyXBs in the Galaxy? This question is closely related to the formation of dormant BHs residing in wide binaries, several of which were recently discovered \citep{Chakrabarti2023,El-Badry2023a,El-Badry2023b,Tanikawa2023,Gaia2024,Wang2024} based on {\tt Gaia} astrometric measurements.

One of the possible reasons for the rarity of BH SyXBs could be related to their low birth rate: the low-mass secondary may not have enough orbital energy to expel the envelope of the massive primary star (i.e., the BH's progenitor) during the common envelope (CE) phase, thus likely leading to a merger of  the binary \citep{Portegies1997,Kalogera1999,Podsiadlowski2003}. However, the known Galactic NS low-mass X-ray binaries outnumber BH low-mass X-ray binaries by a factor of $\sim$2, suggesting that they should have roughly similar birth rates, and this is consistent with theoretical predictions  \citep{Kalogera1998,Podsiadlowski2003,Wang2016MN}. We note that previous studies focused on the formation of low-mass X-ray binaries with relatively short  orbital periods ($\lesssim$ tens of days). Little attention had be paid on the formation of NSs and BHs in wide orbits before the discovery of {\tt Gaia} NSs and BHs.

Additionally, even if there are considerable BH+low-mass giant star binaries in the Galaxy, they might not be luminous X-ray sources. Because of the low accretion rate through wind capture, most of the kinetic energy of the accretion flow is not radiated away but advected into the BH event horizon \citep{Ichimaru1977,Narayan1994}. In addition, a significant fraction of the material captured by the BH may eventually be lost in the form of outflows \citep{Blandford1999}. Consequently, the X-ray emission is too faint to detect \citep{Sen2024}. These qualitative arguments seems reasonable, but systematic quantitative analysis is still lacking.

To address these issues, we employ a BPS method to investigate the formation of BH and NS SyXBs simultaneously. The results can help clarify whether there are a large population of quiescent NSs/BHs with wide-orbit companions in the Galaxy. The rest of the paper is structured as follows. Section 2 describes the method and assumptions.  Our simulated results of the BH SyXB population and observational constraints are displayed in sections 3 and 4, respectively. We summarize the results and discuss several issues in our simulation in section 5.

\section{Method and Assumptions}

We employ the BPS code originally developed by \citet{Hurley2002} and updated by \citet{Shao2014} to simulate the evolution of binary stars. We assume a constant star formation rate of $3\,M_{\odot}$\,yr$^{-1}$ and metallicity of $Z = 0.02$ throughout 12-billion-year lifespan of the Galaxy. The primordial binaries consist of a primary star with mass $M_1$ and a secondary star with mass $M_2$ in circular orbits. The primary star's mass distribution follows the initial mass function (IMF) proposed by \citet{Kroupa1993}, within a mass range from 10 to $60\,M_{\odot}$. The mass ratio ($q = M_2/M_1$) is assumed to be uniformly distributed between 0 and 1 \citep{Kobulnicky2007}. The initial orbital separation ($a$) is drawn from a uniform distribution in the logarithmic scale, ranging from 3 to $10^4$ $R_{\odot}$  \citep{Abt1983}.

We model stellar wind mass loss using the fitting formula in \citet{Nieuwenhuijzen1990}. For OB stars with effective temperatures exceeding 11,000 K and stripped helium stars, we adopt the simulated relations provided by \citet{Vink2001} and \citet{Vink2017}, respectively. We assume that half of the transferred mass from the primary star through Roche-lobe overflow (RLOF) is accreted by the companion star, as suggested by \citet{Shao2014}. The excess material is assumed to be ejected from the binary via isotropic winds, which carry away the specific angular momentum of the accreting star. We use the values of the critical mass ratios ($q_{\rm cr}$) in the simulations by \citet[][see their Table 3]{Iorio2023} to determine the stability of mass transfer. If the mass ratio exceeds $q_{\rm cr}$ the mass transfer becomes dynamically unstable and a CE phase ensues. However, the progenitor systems of SyXBs studied in this work typically consist of a massive ($>10\,M_{\odot}$) primary and a low-mass ($\lesssim 1-2\,M_{\odot}$) secondary, with the mass ratios generally exceeding 5. In such systems, mass transfer through Roche-lobe overflow, if occurs, inevitably leads to CE evolution. Therefore, using other choices of the critical mass ratios \citep[e.g.,][]{Pavlovskii2017,Ge2015,Ge2020,Shao2021} would not significantly affect the results. We utilize the standard energy conservation equation \citep{Webbink1984} to treat CE evolution. In our reference model we set the energy efficiency parameter $\alpha = 1.0$. 
The binding energy parameter $\lambda$ of the stellar envelope is dependent on the star's mass and evolutionary state \citep{Ivanova2013}. Based on the work of \citet{Xu2010a,Xu2010b}, \citet{Wang2016RAA} employed the  stellar evolution code \texttt{MESA} \citep{Paxton2011,Paxton2013,Paxton2015} to systematically calculate the $\lambda$ values, adopting modified definition of the stellar core-envelope boundary. Here we follow the method of \citet{Wang2016RAA} to conduct calculations of the $\lambda$ values with \texttt{MESA} for denser grids of stellar masses and radius. 

Both NSs and BHs are evolutionary products of massive stars, but it is still in debate which stars produce NSs and which stars produce BHs \citep[][for a recent review]{Boccioli2024}. Considering the big uncertainties in the mechanisms of supernova (SN) explosions, we construct four SN models to account for both the remnant masses and the imparted natal kicks to the newborn NSs/BHs \citep[see also][]{Deng2024a}:
\begin{enumerate}
    \item Rapid explosion model (SN A): The SN explosions occur within the first 250 ms after the bounce. This model, as described in \citet{Fryer2012}, determines the compact object's mass based on the core's mass at the explosion time. It should be noted that the SN A model is incapable of producing compact objects with masses in the $2-5\,M_{\odot}$ range.
    \item Delayed explosion model (SN B): The SN explosions  occur over a much longer timescale than 250 ms. Similar to the rapid model \citep{Fryer2012}, this model uses the core mass at the explosion to determine the final mass, with a continuous distribution of the compact object mass.
    \item Stochastic explosion model (SN C): This is a probabilistic recipe for SN explosions that parameterize the remnant mass and kick as a function  of the CO core mass and the He shell mass of the progenitor star \citep{Mandel2020}.
    \item Failed explosion model (SN D): Based on \citet{O'Connor2011}, this model assumes that the core's compactness at the collapse dictates its fate. The cores with low compactness undergo a successful SN and form NSs, while highly compact cores experience a failed explosion, resulting in BHs. The BH mass in this model is directly related to the pre-SN helium or CO core mass \citep[e.g.,][]{Wang2016MN}.
\end{enumerate}
Besides CCSNe, stars with initial mass
$\sim 9-12\,M_\sun$ are thought to produce electron-capture supernovae (ECSNe). However, there are considerable uncertainties in the
triggering condition of ECSNe \citep{Nomoto1984,Podsiadlowski2004,Woosley2015,Doherty2017,Poelarends2017}. Here we assume that if the helium core mass is the range of $1.83-2.25\,M_\sun$, the star may
eventually explode in an ECSN \citep{Shao2018a}. Note that the formation of BHs does not depend on the choice of the ECSN condition.

In line with \citet{Fryer2012}, we assume that the gravitational mass of the remnant is $90\%$ of its baryonic mass. For all models, any compact object exceeding $3\,M_{\odot}$ is regarded to be a BH. We do not consider BHs formed from accretion-induced collapse of NSs. 

For BHs originating from rapid, delayed, and failed SN models, we assume that their kick velocities scale inversely with their masses. This relationship is expressed as $V_{\rm kick,BH}=(3\,M_{\odot}/M_{\rm BH})V_{\rm kick,NS}$, where $V_{\rm kick,NS}$ represents the kick velocity imparted to newborn NSs. In the stochastic SN model, both the BH mass and kick velocity follow a distinct probability distribution, determined by the pre-explosion CO core mass \citep{Mandel2020}. We further differentiate between the NS kick velocities in different types of SNe: CCSN NS kicks follow a Maxwell distribution with the dispersion velocity $\sigma_{\rm CCSN}=$ 265 km\,s$^{-1}$ \citep{Hobbs2005}, and ECSN NS kicks have a lower dispersion velocity $\sigma_{\rm ECSN}=$ 30 km\,s$^{-1}$ \citep{Podsiadlowski2004,Heuvel2004,Verbunt2017,Deng2024a}. 

After their formation, the compact objects would interact with the stellar winds from the companions. We employ the Bondi–Hoyle–Lyttleton (BHL) accretion model \citep{Hoyle1939,Bondi1944} to estimate the accretion rate:
\begin{equation}
    \dot{M}_{\rm acc} = \frac{1}{4}\frac{1}{\sqrt{1-e^2}} \dot{M}_{\rm wind}\left(\frac{v_{\rm rel}}{v_{\rm w}}\right)\left(\frac{R_{\rm acc}}{a}\right)^2,
\end{equation}
where $\dot{M}_{\rm wind}$ is the mass loss rate of the giant star, calculated according to the fitting formula of \citet{Nieuwenhuijzen1990}, $v_{\rm w}$ is the wind velocity, $v_{\rm rel}^2=v_{\rm w}^2+v^2_{\rm orb}$ is the relative stellar wind velocity with $v_{\rm orb}$ being the orbital velocity, and $R_{\rm acc}$ is the accretion radius of the compact object. To evaluate the impact of wind velocities on the calculated results, we consider two wind velocity values: $v_{\rm w}=0.25 v_{\rm esc}$ and $v_{\rm esc}$ for a given mass loss rate\footnote{The wind density then varies inversely with $v_{\rm w}$ in the case of spherically expanding wind.}, where $v_{\rm esc}$ is the escape velocity from the giant star's surface.  The accretion radius 
\begin{equation}
R_{\rm acc}=\delta \frac{2GM_{\rm CO}}{v^2_{\rm rel}},
\end{equation}
where $G$ is the gravitational constant, and $M_{\rm CO}$ is the compact object mass. The parameter $\delta$ was introduced by \citet{Hunt1971} to account for the bow shock location within a stellar wind. While \citet{Liu2017} suggested a $\delta$ value between 0.3 and 0.5, other studies \citep[e.g.,][]{deVal2017} argued for a potential increase of $\dot{M}_{\rm acc}$ due to gravitational focusing.  Therefore, we set $\delta=0.4$ and 1.

Note that $R_{\rm acc}$ should be smaller than the Roche lobe radius $R_{\rm L,1}$ of the compact object in a binary system. Thus, we let
\begin{equation}
    R_{\rm acc}={\rm min}(\delta \frac{2GM_{\rm CO}}{v^2_{\rm rel}},R_{\rm L,1}).
\end{equation}

In the case of wind accretion by NSs, the NS spin and magnetic field also play an important role in modulating the accretion rate. To account for these factors, we initialize the NS spin and magnetic field based on the population study of pulsars by \citet{Faucher2006} and simulate the magnetic field decay caused by accretion as in \citet{Kiel2008} and \citet{Oslowski2011}. To determine whether a wind-fed NS is in the accretor state and appears as an X-ray source, we adopt the criterion proposed by \citet[][see their Table 1]{Lv2012}. If the X-ray luminosity is $<4\times 10^{36}$ erg\,s$^{-1}$, the NS resides the subsonic settling accretion regimes \citep{Shakura2018}, and we utilize the model in \citet{Yungelson2019} to calculate the accretion rate and the spin evolution of the NS. 

In our simulations, we identify SyXBs based on the following criteria:
\begin{enumerate}
    \item The companion star is in the first giant branch (FGB), core He-burning giant branch (CHeB), or early asymptotic giant branch (EAGB) phase;
    \item The binary is detached ($R_2/R_{\rm L,2}<0.8$);
    \item The accretion rate $\dot{M}_{\rm acc}$ of the compact object is greater than $10^{-15}$ $M_{\odot}$ yr$^{-1}$.
\end{enumerate}
Our results presented in the following figures and tables refer to binaries that simultaneously meet all of these three conditions.

We construct 16 models considering various combinations of the initial parameters, and evolve 10$^8$ binary systems in each model.
Table 1 lists the model parameters and the predicted numbers of BH and NS SyXBs in different models. We see that the SN A and SN B models hardly produce any BH SyXBs. The main reason is that the primary star (the BH's progenitor star) in these two models is so massive ($>20\,M_\sun$) that the low-mass companion star does not have enough orbital energy to expel the star's envelope during CE evolution, leading to binary merger instead \citep[see also][]{Podsiadlowski2003,Wang2016MN,Shao2019,Deng2024b}. In the SN C model a substantial number ($\sim 70-150$, depending on the wind velocity and the accretion radius) of BH SyXBs can form, while the SN D model yields only a few ($\sim 5-7$) BH SyXBs. All the four SN models produce a much larger NS SyXB population (with the total number $\sim 300-2000$), which are insensitive to the SN models. The changes in $v_{\rm w}$  and $\delta$ lead to the variation in the number of BH SyXBs by a few times. Overall, the number of NS SyXBs exceeds that of BH SyXBs by at least a factor of 6.

\begin{deluxetable*}{lcccccc}
\tablenum{1}
\tablecaption{Model parameters and simulated results}
\tabletypesize{\small}
\tablehead{
	\multirow{2}{*}{Model}             & \multirow{2}{*}{CCSN}   & $v_{\rm{w}}$ & Kick velocity$^{\#}$  & \multicolumn{2}{c}{Numbers} &\multirow{2}{*}{$N_{\rm{NS\,SyXBs}}/N_{\rm{BH\,SyXBs}}$} \\
  &        & ($v_{\rm esc}$)  &   $\sigma_{\rm BH},\sigma_{\rm NS, CCSN}$ & BH SyXBs  & NS SyXBs &  }
\startdata
\hline
\multicolumn{7}{c}{$R_{\rm acc}=2GM/v_{\rm rel}^2$}\\
\hline
	SNA0.25V$_{\rm esc}$1R$_{\rm acc}$  & rapid       & 0.25 &    (3/$M_{\rm BH}$)265,\,265   & $\ll 1$  &  1895 & -  \\
	SNA1V$_{\rm esc}$1R$_{\rm acc}$     & rapid       & 1 &      (3/$M_{\rm BH}$)265,\,265    & $\ll 1$  &  840 & - \\
	SNB0.25V$_{\rm esc}$1R$_{\rm acc}$  & delay       & 0.25 &   (3/$M_{\rm BH}$)265,\,265    & $\ll 1$  &  1889 & - \\
	SNB1V$_{\rm esc}$1R$_{\rm acc}$     & delay       & 1 &      (3/$M_{\rm BH}$)265,\,265    & $\ll 1$  &  831 & - \\
	SNC0.25V$_{\rm esc}$1R$_{\rm acc}$  & stochastic  & 0.25 &     $^*$                       & 151 &  2148 & 14 \\
    SNC1V$_{\rm esc}$1R$_{\rm acc}$     & stochastic  & 1 &     $^*$                          & 89  &  914 & 10 \\
    SND0.25V$_{\rm esc}$1R$_{\rm acc}$  & failed      & 0.25 &   (3/$M_{\rm BH}$)265,\,265    &  7  &  1914 & 273 \\
	SND1V$_{\rm esc}$1R$_{\rm acc}$     & failed      & 1 &      (3/$M_{\rm BH}$)265,\,265    &  5  &  808 & 162 \\
\hline
\multicolumn{7}{c}{$R_{\rm acc}=0.4\times 2GM/v_{\rm rel}^2$}\\
\hline
    SNA0.25V$_{\rm esc}$0.4R$_{\rm acc}$  & rapid       & 0.25 &    (3/$M_{\rm BH}$)265,\,265   & $\ll 1$ & 927 & - \\
    SNA1V$_{\rm esc}$0.4R$_{\rm acc}$     & rapid       & 1 &       (3/$M_{\rm BH}$)265,\,265   & $\ll 1$ & 358  & -  \\
    SNB0.25V$_{\rm esc}$0.4R$_{\rm acc}$  & delay       & 0.25 &    (3/$M_{\rm BH}$)265,\,265   & $\ll 1$ & 912  & -    \\
    SNB1V$_{\rm esc}$0.4R$_{\rm acc}$     & delay       & 1 &       (3/$M_{\rm BH}$)265,\,265   & $\ll 1$ & 352  & - \\
    SNC0.25V$_{\rm esc}$0.4R$_{\rm acc}$  & stochastic  & 0.25 &       $^*$                      & 88 &  1036 & 12 \\
    SNC1V$_{\rm esc}$0.4R$_{\rm acc}$     & stochastic  & 1 &       $^*$                         & 70 &  395   & 6   \\
    SND0.25V$_{\rm esc}$0.4R$_{\rm acc}$  & failed      & 0.25 &   (3/$M_{\rm BH}$)265,\,265    & 5  &  886  & 177   \\
    SND1V$_{\rm esc}$0.4R$_{\rm acc}$     & failed      & 1 &      (3/$M_{\rm BH}$)265,\,265    & 5  &  335 & 67  \\
\enddata
\tablecomments{\\
$^{\#}$ NSs born in ECSN receive a kick with $\sigma_{\rm NS,ECSN}=30$ km\,s$^{-1}$. \\
$^*$ Satisfy specific probability distributions described in \citet{Mandel2020}.
}
\end{deluxetable*}

\section{Potential Symbiotic X-ray binary Population in the Galaxy}

Figures 1 compares the distributions of the companion mass and the orbital period for the simulated BH and NS SyXBs (left and right panels respectively). The colors in each panel denote the number distribution of the binaries.  The low-right boundary in each subplot is determined by the criteria we use to identify SyXBs. Here we adopt the wind velocity of $v_{\rm w}=0.25\,v_{\rm esc}$ and  $\delta=1$. As seen from the left panel, there are very few BH SyBs in both the SN A and SN B models. In the SN C and SN D models, the BH's progenitor star can be substantially less massive than in the SN A and SN B models, thus allowing the formation of a considerable number of BH SyXBs. It is worth noting that in the SN C model the orbital periods of BH SyXBs show bimodal distribution, clustered around $10^2-10^3$ days and close to $10^4$ days. The latter results from the evolutionary routes without the CE phase. In comparison, the four SN models exhibit similar distributions of NS SyXBs. Since the NS progenitors are generally less massive than the BH progenitors, more NS binaries can survive CE evolution.  In addition, more than $80\%$ NSs in our simulations are born in ECSNe with relatively low kick velocities.

\begin{figure}
    \centering
    \plottwo{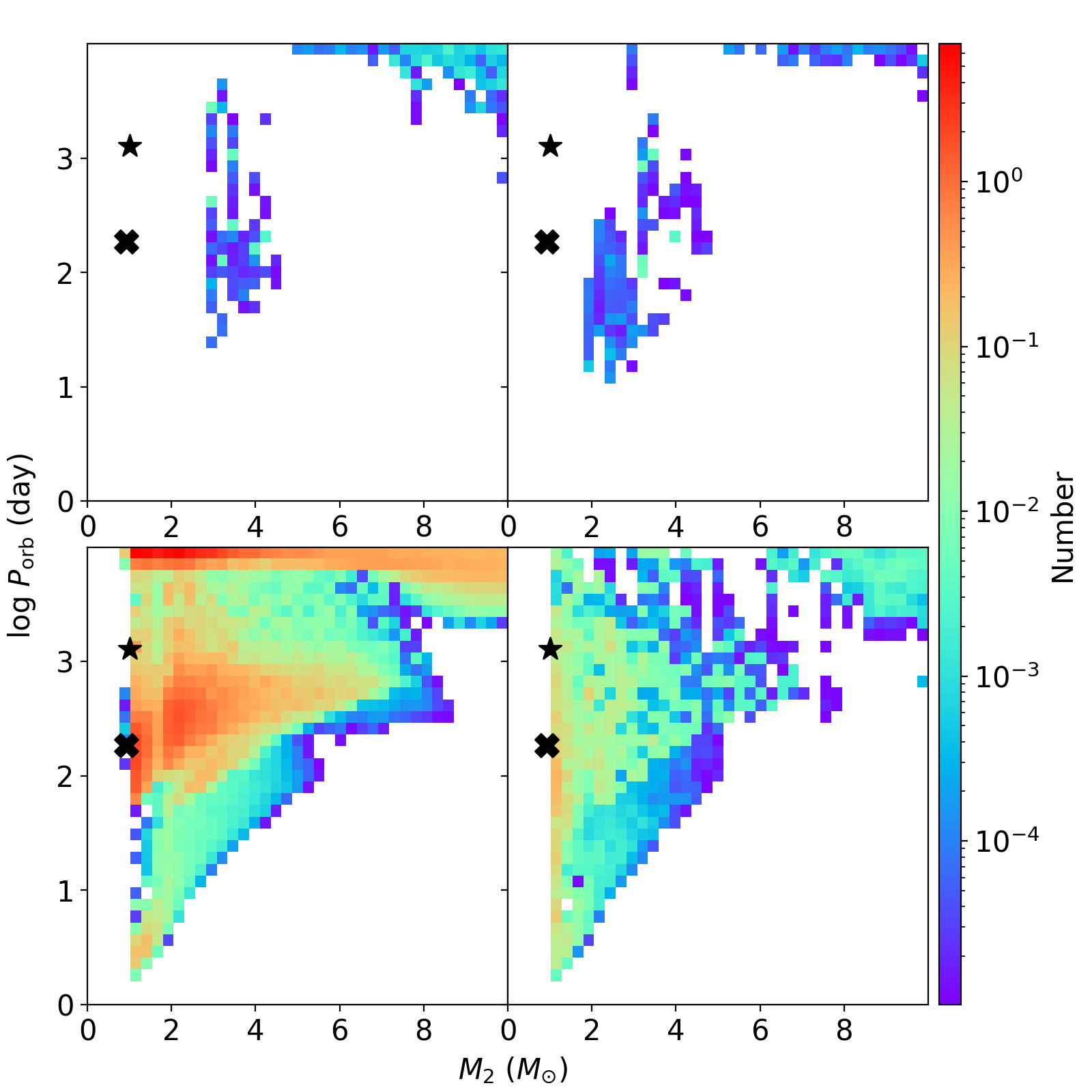}{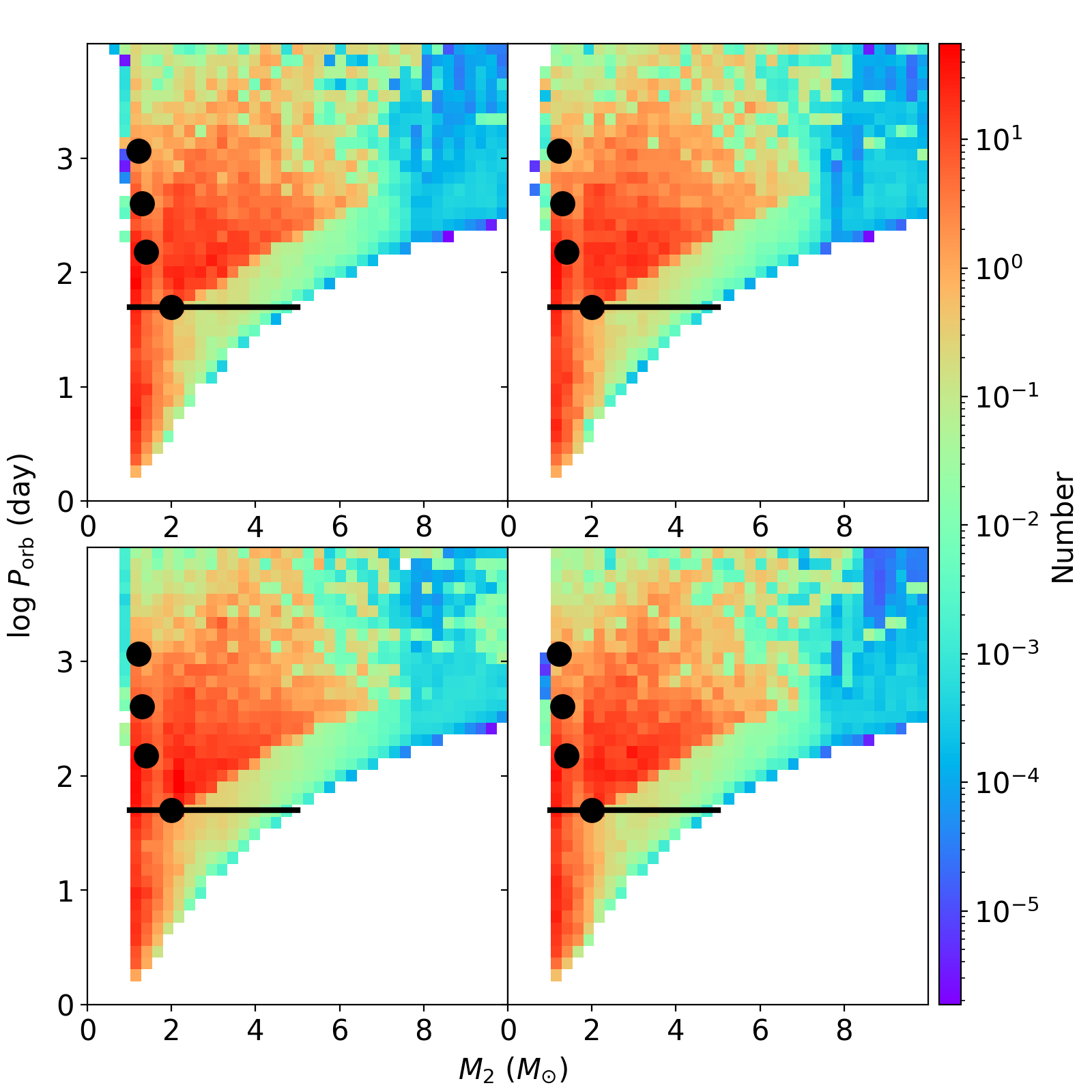}
    \caption{Distribution of companion masses and orbital periods for BH (left) and NS (right) SyXBs in different SN models. In all models we adopt $v_{\rm w}=0.25\,v_{\rm esc}$ and $\delta=1$. The cross and star represent Gaia BH1 and Gaia BH2 \citep{El-Badry2023a,El-Badry2023b}, respectively. The black dots represent NS SyXBs with the companion mass and orbital period measured \citep[data taken from][]{Masetti2002,Hinkle2006,Nespoli2010,Masetti2011}.}
    \label{fig:1}
\end{figure}

To see the properties of the simulated BH SyXBs in detail, we select the model results in SNC0.25V$_{\rm esc}$1R$_{\rm acc}$ and SND0.25V$_{\rm esc}$1R$_{\rm acc}$, which are most productive for BH SyXBs, and display the distributions and the frequency profiles of the companion mass, BH mass, orbital period, eccentricity, and X-ray luminosity\footnote{Here we simply assume the X-ray luminosity $L_{\rm X}=(1/12)\dot{M}_{\rm acc}c^2$.} in Figure 2.
The companion masses are primarily distributed $\sim 1-4\,M_{\odot}$ (over $80\%$ of the systems have companions less massive than $3\,M_{\odot}$). The BH masses exhibit a bimodal distribution peaked at $\sim 4\,M_{\odot}$ and $\sim 7\,M_{\odot}$ in the SNC model and are primarily distributed around $4-5\,M_{\odot}$ in the SND model. More than half of the systems have eccentricities less than 0.2 in the SN C model, indicating that most of the BHs are born with small kicks in this case. However, the eccentricity distribution is much wider in the SN D model, because we set the BH kick velocity to be the NS kick velocity reduced by a factor of the BN/NS mass ratio.  The gap in the eccentricity vs. orbital period plots reflects whether or not the formation process involved CE evolution. The orbital period distributions indicate that most of the component stars in the SN C model evolve more or less independently, while most BH SyXBs in the SN D model have experienced CE evolution. In both models the X-ray luminosities are distributed within the range of $\sim 10^{31}-10^{37}$ erg\,s$^{-1}$. 

Recent Gaia observations have detected a few detached BH binaries  with a low-mass secondary. The cross and star in Figure 2 represent Gaia BH1 and BH2 \citep{El-Badry2023a,El-Badry2023b}, respectively. Gaia BH1 consists of a $\sim 9.62\,M_{\sun}$ BH and a $0.93\,M_{\sun}$ dwarf star in a 185.6 day orbit with an eccentricity of 0.45, and Gaia BH2 consists of a $\sim 8.9\,M_{\sun}$ BH and a $\sim 1\,M_{\sun}$ red giant star in a 1277 day orbit with an eccentricity of 0.52. Both sources are lacking  detectable X-ray emission, but are expected to evolve to be BH SyXBs when the secondary stars become red giants. Our simulations can reproduce the companion mass, orbital period, and eccentricity of these two binaries, but fail to account for the BH masses, because very massive progenitor stars are required to produce $\sim 9\,M_\sun$ BHs.  \citet{El-Badry2023b} suggested that the BHs could have evolved from stars exceeding $65\,M_{\odot}$, losing their hydrogen envelope in the late main sequence without undergoing CE. Our study did not consider this pathway, primarily due to the big uncertainties in the mass-loss efficiency of massive stars. There are alternative proposals that the binaries could be formed from dynamic processes \citep{Di2024} or from three-body interactions \citep{El-Badry2023b,Di2024}.

\begin{figure}
    \centering
    \includegraphics[width=0.9\textwidth]{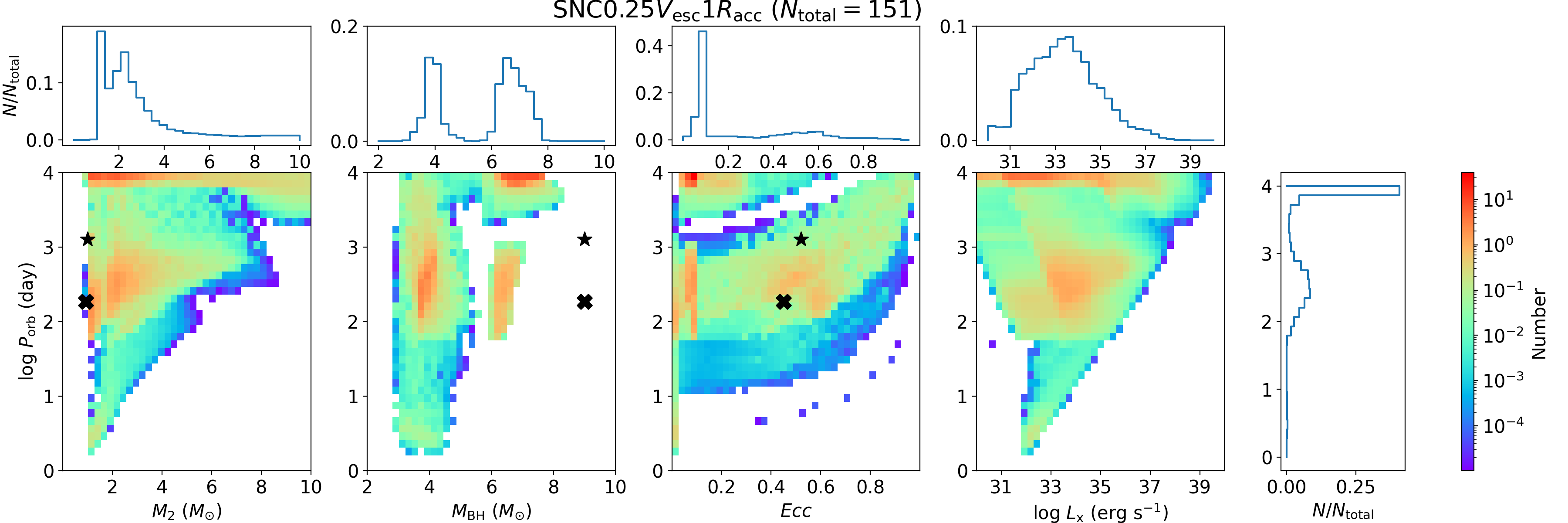}   \includegraphics[width=0.9\textwidth]{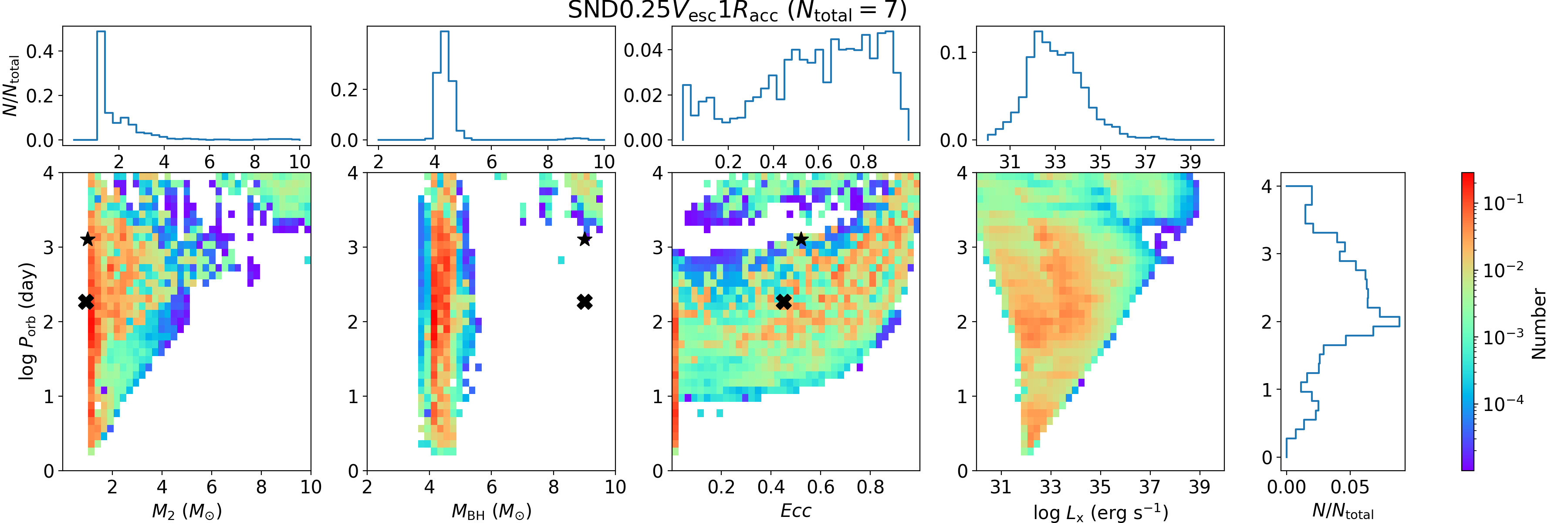}
    \caption{Distributions of BH SyXBs in the companion mass vs. orbital period, BH mass vs. orbital period, eccentricity vs. orbital period, and X-ray luminosity vs. orbital period diagrams (from left to right) in the SNC0.25V$_{\rm esc}$1R$_{\rm acc}$ (upper panels) and SND0.25V$_{\rm esc}$1R$_{\rm acc}$ (lower panels) models. The cross and star represent Gaia BH1 and Gaia BH2 \citep{El-Badry2023a,El-Badry2023b}, respectively.}
    \label{fig:2}
\end{figure}

\section{Detectable BH symbiotic X-ray binary Population in the Galaxy}

The results in Section 3 demonstrate that both the SN C and SN D models can produce BH SyXBs through isolated binary evolution, but the overall number of BH SyXBs are significantly smaller than that of NS SyXBs. In this section,  we discuss the detectable number of BH SyXBs and their parameter distributions. 

A wind-fed BH can be detectable in X-ray only when its flux is above the detection limit. This requires that the captured material by the BH has sufficient angular momentum $j$ to form an accretion disk, since spherical accretion onto BHs is highly inefficient in X-ray radiation \citep{Shapiro1976}. In the framework of BHL accretion, the specific angular momentum of the captured wind material is
\begin{equation}
    j=\eta \Omega_{\rm orb}R^{2}_{\rm acc},
\end{equation}
where $\Omega_{\rm orb}$ is the orbital angular velocity and $\eta= 0.25$ according to \citet{Illarionov1975}. The corresponding circularization radius is
\begin{equation}
    R_{\rm CRIC}=\frac{j^2}{GM_{\rm BH}}.
\end{equation}
Disk formation around a NS or BH requires
\begin{equation}
    R_{\rm CRIC}>R_{\rm m} \ {\rm or}\ R_{\rm ISCO},
\end{equation}
where $R_m$ is the NS magnetospheric radius approximated by the Alfv\'en radius
\begin{equation}
    R_{\rm m}\simeq R_{\rm A}=\left(\frac{\mu^2}{\dot{M}_{\rm acc}\sqrt{2GM_{\rm NS}}}\right)^{2/7},
\end{equation}
where $\mu$ and $M_{\rm NS}$ are the magnetic moment and mass of the NS,
and $R_{\rm ISCO}$ is the radius of the innermost stable orbit for a Schwarzchild BH,
\begin{equation}
    R_{\rm ISCO}=3R_{\rm Sch}=\frac{6GM_{\rm BH}}{c^2},
\end{equation}
where $R_{\rm Sch}$ is the Schwarzchild radius. 

The X-ray radiation efficiency of a BH accretion disk depends on the structure of the disk and possible mass loss from the disk. At sufficiently low accretion rate ($\lesssim 10^{-2}\dot{M}_{\rm Edd}$, where $\dot{M}_{\rm Edd}$ is the Eddington accretion rate), radiative cooling in the inner region of the disk becomes inefficient, forming advection-dominated accretion flow
(ADAF) in the inner disk \citep{Narayan1994,Narayan1995}. Meanwhile, numerical simulations observe the presence of outflows in ADAF \citep[e.g.,][]{Stone1999,Yuan2010}. The mass flow rate is assumed to scale with the radius from the BH following a power law $\dot{M}(R)\propto (R/R_{\rm ADAF})^s$ \citep[][]{Blandford1999}, where $R_{\rm ADAF}$ is the transition radius between the inner ADAF and outer thin disk\footnote{For disk-accreting NSs, if $R_{\rm m}<R_{\rm ADAF}$, we also consider disk wind loss and similar law for the mass flow rate. Otherwise, the disk is assumed to be optically thick and geometrically thin.}. Then the BH accretion rate is given by
\begin{equation}
    \dot{M}_{\rm ISCO}=\dot{M}_{\rm acc}\left(\frac{R_{\rm ISCO}}{R_{\rm ADAF}}\right)^s.
\end{equation}
The transition radius can be empirically determined from the observations of transient X-ray binaries and supermassive BHs in other galaxies \citep{Yuan2004,Cao2014},
\begin{equation}
R_{\rm ADAF}\simeq R_{\rm Sch}\left(\frac{\dot{M}_{\rm acc}}{\dot{M}_{\rm Edd}}\right)^{-0.5}.
\end{equation}
The power index $s$ is set to be 0.4 following \citet{Xie2012} and \citet{Yuan2014}.

The accretion luminosity is given by
\begin{equation}
    L_X=\epsilon \dot{M}_{\rm ISCO}c^2.
\end{equation}
We employ the fitting formula provided by \citet{Xie2012} to calculate the radiation efficiency for ADAF,
\begin{equation}  \epsilon=\epsilon_0\left(\frac{\dot{M}_{\rm ISCO}}{\dot{M}_{\rm c}}\right)^{a},
\end{equation}
where $\epsilon_0$ and $a$ are taken from Table 1 of \citet{Xie2012}, and $\dot{M}_{\rm c}=10^{-2}\dot{M}_{\rm Edd}$. 

Figure 3 compares different characteristic radii as a function of the wind velocity for BH and NS SyXBs with typical parameters. It is seen that, for BH binaries, $R_{\rm ISCO}$ is always smaller than both $R_{\rm CIRC}$ and $R_{\rm ADAF}$, indicating the formation of accretion disks and the presence of ADAF. Our simulations show that there are accretion disks in over $90\%$ of BH SyXBs. In comparison, for NSs, the relative $R_{\rm m}$ and $R_{\rm CIRC}$ can vary depending on the wind velocity, accretion radius, and accretion rate, indicating that accretion onto NSs can be either disk accretion or spherically symmetric wind accretion. Additionally, $R_{\rm ADAF}$ is always smaller than $R_{\rm m}$ for NSs, suggesting that ADAF does not occur in the disk accretion case. 

Given the X-ray luminosity, we can estimate the flux according to the distance between the X-ray source and the Sun. Following \citet{Faucher2006}, we adopt a spatial distribution for BH SyXBs in the Milky Way as follows,
\begin{equation}
    p(r;{\rm Gauss})\propto \exp[-\frac{(r-R_{\rm peak})^2}{2\sigma^2_{r}}],
\end{equation}
where $r$ is the BH's distance from the Galactic center, $R_{\rm peak}=7.04$ kpc and $\sigma_{\rm r}=1.83$ kpc. For each SyXB, we calculate its distance from the Sun and obtain the flux density. We set the flux density limit for SyXB detection to be $F_{\rm lim}\simeq 10^{-11} \rm erg \,s^{-1} cm^{-2}$ \citep{Sen2021}.

Table 2 presents the calculated numbers of detectable BH and NS SyXBs in the Galaxy and their ratios under several representative models. We can see that the number ratios become larger compared with those in Table 1, strengthening our former conclusion that NS SyXBs significantly overnumber BH SyXBs.
Figure 4 shows the distributions of detectable BH SyXBs in the SN C and SN D models with the same input parameters as in Figure 2. The BH SyXB numbers decease by roughly an order of magnitude, to $\sim 8-18$ and $\sim 0.18-0.45$ in the SNC and SND models respectively. While the orbital periods and eccentricities follow similar distributions as in Figure 2, the companion masses have a much wide distribution, with the majority above $4\,M_{\odot}$. The X-ray luminosities also enhance to $\sim 10^{35}-10^{37}$ erg\,s$^{-1}$.  These discrepancies stem from the fact the X-ray luminosities scale with the wind loss rates non-linearly as revealed by Eqs. (10)-(11). Thus, binaries with lower-mass companions are too faint to be detectable.   

\begin{figure}
    \centering
    \includegraphics[width=0.8\linewidth]{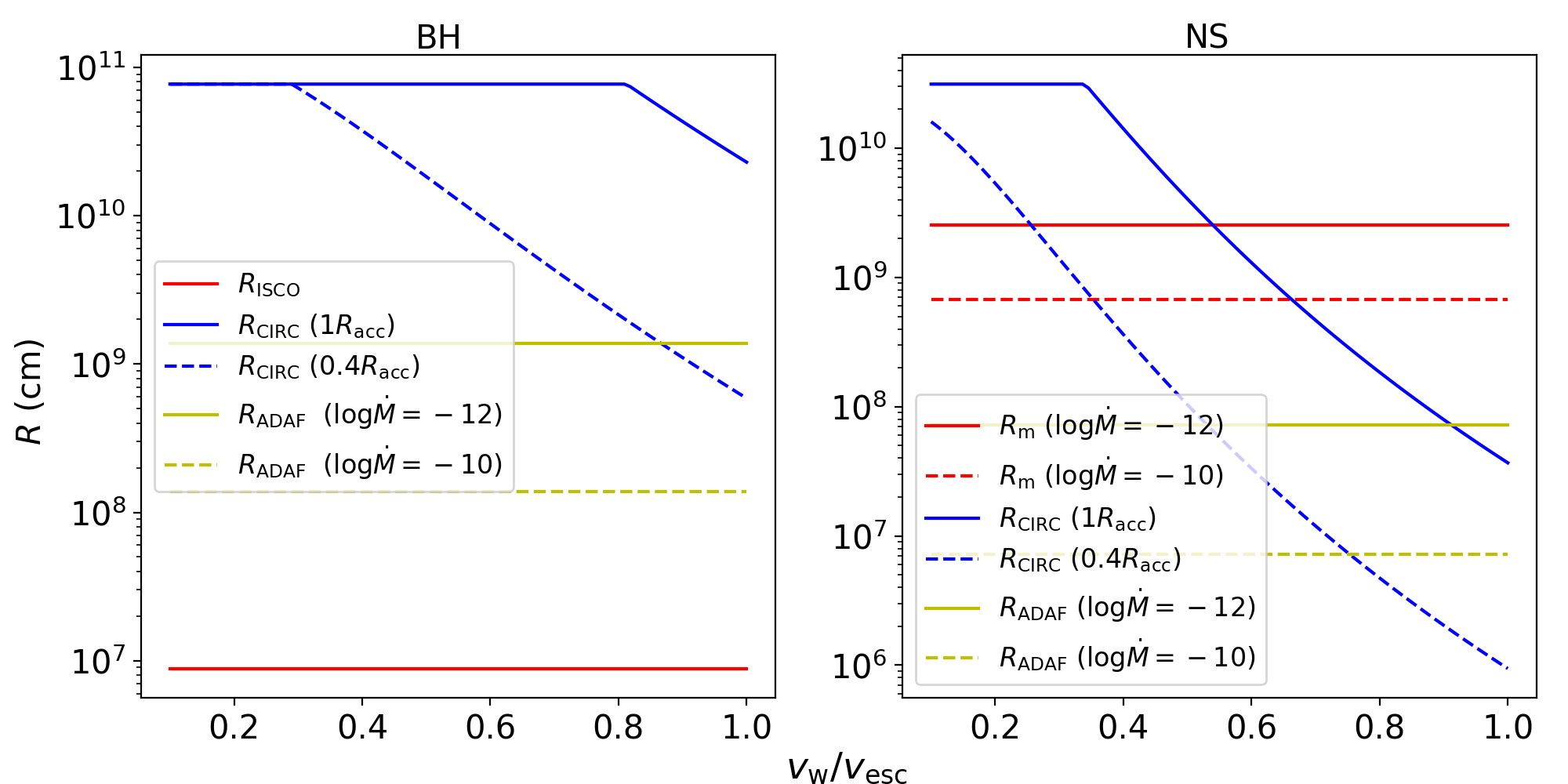}
    \caption{The characteristic radii $R_{\rm ISCO}$, $R_{\rm m}$ (red), $R_{\rm CIRC}$ (blue), and $R_{\rm ADAF}$ (yellow) for BH (left) and NS (right) SyXBs. We adopt the BH mass of 10 $M_{\sun}$, NS mass of 1.4 $M_{\sun}$, companion mass of 2 $M_{\sun}$ and radius of 10 $R_{\sun}$,   orbital separation of 200 $R_{\sun}$, and NS surface magnetic field of $10^{12}$ G. The accretion rates are set to be $10^{-12}\,M_{\odot}\,$yr$^{-1}$ and $10^{-10}\,M_{\odot}\,$yr$^{-1}$ for comparison.}
    \label{fig:3}
\end{figure}

\begin{figure}
    \centering
    \includegraphics[width=0.9\textwidth]{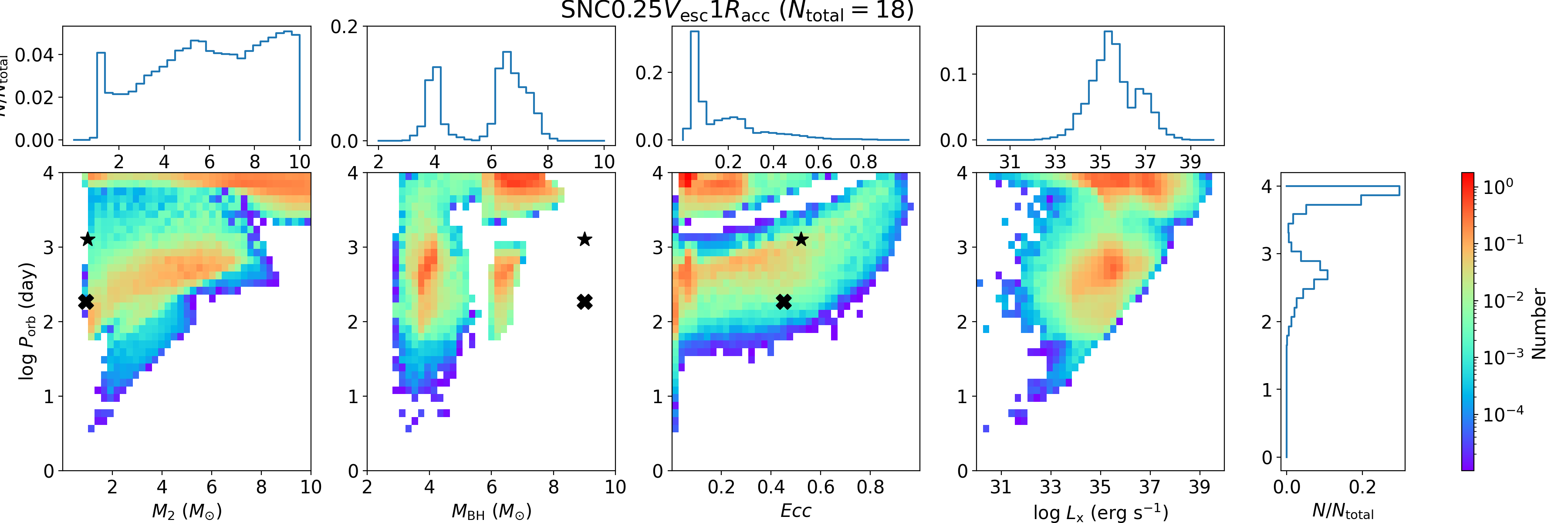}
    \includegraphics[width=0.9\textwidth]{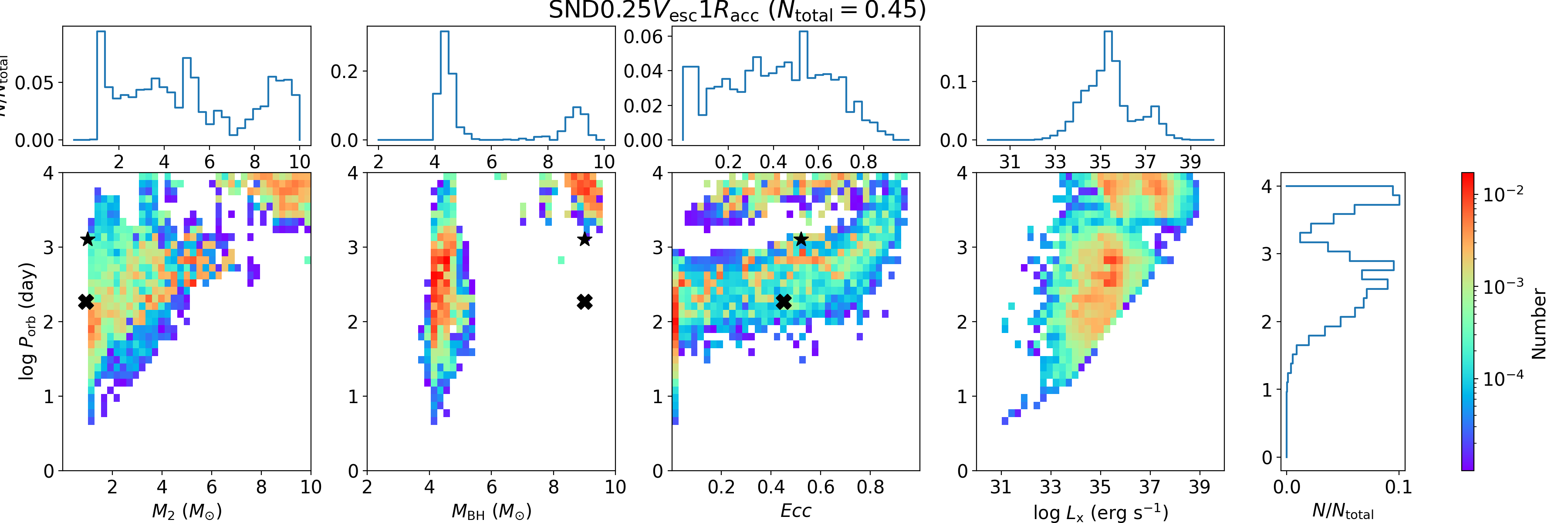}
    \caption{Similar to Figure 2, but for detectable BH SyXBs.}
    \label{fig:4}
\end{figure}

Figure 5 presents the probability density function (PDF) of the SyXB lifetimes in the SNC0.25V$_{\rm esc}$1R$_{\rm acc}$ model. BH SyXBs and NS SyXBs predominantly exhibit lifetimes around $10^4-$a few $10^8$ yrs and $10^2-$a few $10^8$ yrs, respectively. 
The lifetime reduction seems to be more significant for BH SyXBs than NS SyXBs, indicating that the scarcity of detectable BH SyXBs is also partly caused by the change in the duration of SyXBs.

\begin{deluxetable*}{cccc}
\tablenum{2}
\tablecaption{Detectable SyXBs Numbers}
\tabletypesize{\small}
\tablehead{
	\multirow{2}{*}{Model}           & \multicolumn{2}{c}{Detectable Numbers} &\multirow{2}{*}{$N_{\rm{NS\,SyXBs}}/N_{\rm{BH\,SyXBs}}$} \\
  &        BH SyXBs  & NS SyXBs &  }
\startdata
\hline
	SNC0.25V$_{\rm esc}$1R$_{\rm acc}$     & 8-18 &  336 & 19-42 \\
    SNC1V$_{\rm esc}$1R$_{\rm acc}$      &  1.9-5.2 &  41 & 8-22 \\
    SND0.25V$_{\rm esc}$1R$_{\rm acc}$   &  0.18-0.45  &  309 & 687-1883 \\
	SND1V$_{\rm esc}$1R$_{\rm acc}$        &  0.05-0.14  &  35 & 250-700 \\
    SNC0.25V$_{\rm esc}$0.4R$_{\rm acc}$    & 2.5-6.1&  36 & 6-14 \\
    SNC1V$_{\rm esc}$0.4R$_{\rm acc}$       & 0.3-1.0 &  6   & 6-20   \\
    SND0.25V$_{\rm esc}$0.4R$_{\rm acc}$    & 0.06-0.17  &  32  & 188-533   \\
    SND1V$_{\rm esc}$0.4R$_{\rm acc}$       & $\ll 1$  &  5 & - \\
\enddata
\end{deluxetable*}

\begin{figure}
    \centering
    \includegraphics[width=0.8\linewidth]{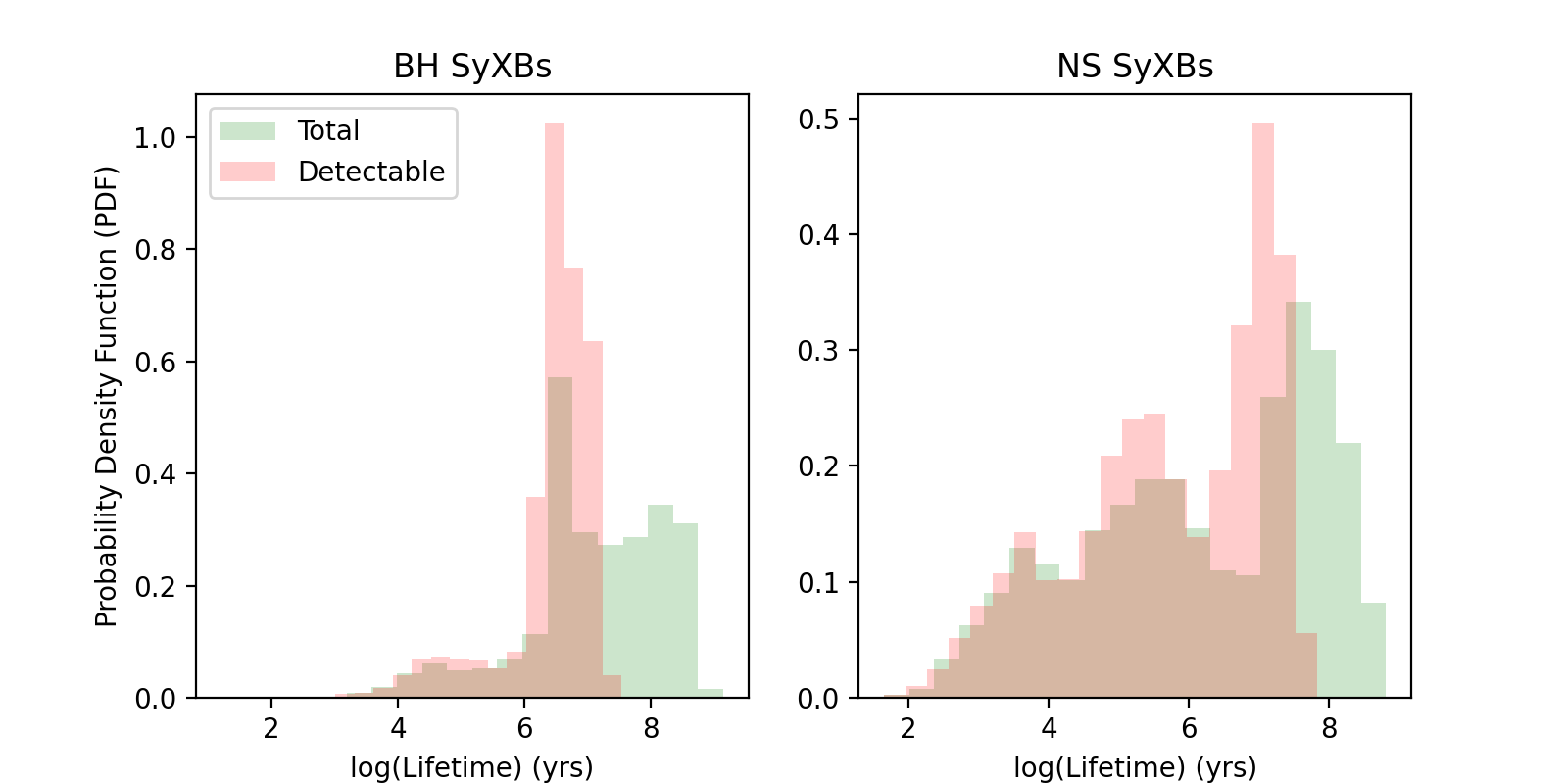}
    \caption{The left and right panels display the lifetime PDFs for BH and NS SyXBs, respectively. The green and red bars represent the total and detectable lifetimes of the SyXBs, respectively. }
    \label{fig:5}
\end{figure}

\section{Conclusions and Discussion}

To date, over a dozen NS SyXBs have been identified in the Galaxy, but there are no BH SyXBs detected. Given the fact that traditional NS and BH low-mass X-ray binaries have roughly similar birthrates \citep[e.g.,][]{Kalogera1998,Podsiadlowski2003,Wang2016MN}, it is natural to ask why BH SyXBs have remained elusive. This study aims to address this question. 

We employ various SN models and kick velocities to investigate the formation of NS and BH SyXBs. Our simulations reveal that while the SN A and SN B models are unable to produce any BH SyXBs, the SN C and SN D models can generate $70-150$ and $5-7$ BH SyXBs in the Galaxy, respectively. However, even in the most optimistic SNC0.25V$_{\rm esc}$1R$_{\rm acc}$ model, only around one tenth of them ($\sim 10$) can be detected as X-ray sources. In comparison, there are potentially $\sim 300-2000$ NS SyXBs. This disparity stems from two facts. First,
the probability of successfully forming  wide-orbit BH+low-mass secondary star binaries is significantly less than that of NS+low-mass secondary star binaries, not only because there are fewer BH progenitor stars (which are generally more massive than the NS progenitor stars) due to the IMF, but also because the majority of the primordial binaries with very large mass ratio are more likely to merge during CE evolution.
Second, most BH SyXBs, although have similar accretion rates as NS SyXBs, are likely to be much more inefficient in producing X-ray radiation, because the inner accretion disks are likely in the form of an ADAF.    

Our results are subject to several issues. The first is the SN mechanisms. In the SN A and B models, stars with initial masses  $\gtrsim 20\,M_\sun$ are expected to form BHs \citep{Wang2016MN}.  \citet{Smartt2009} showed that there is a lack of detected high-luminosity progenitor stars of type IIp SNe, suggesting that red supergiants with mass $\gtrsim 17\,M_\sun$ do not produce CCSNe \citep[see also,][for discussion]{Jerkstrand2015J,Davies2020}. One potential solution is that the cores of such red supergiants may collapse directly to form a BH without an enough strong shock to explode the star entirely \citep[e.g.,][]{Kochanek2008}. There have been efforts trying to search for such so-called failed SNe, and a few candidates have been discovered \citep{Gerke2015,Reynolds2015,Adams2017,Basinger2021,Neustadt2021,Kochanek2024}. Theoretically, recent 3D simulations of CCSNe showed that the explodability of the cores varies non-monotonically with its initial mass
\citep{O'Connor2011,Pejcha2015,Ertl2016,Muller2016,Sukhbold2016,Couch2020,Burrows2023,Burrows2024}. In both the SN C and SN D models the BH progenitor stars can be less than $20\,M_\sun$. Thus, the detection of BH SyXBs could provide useful evidence to test different SN models.

CE evolution also plays an important role in the formation of BH SyXBs. To investigate the impact of the CE parameter $\alpha$ on our results, we calculate the total and detectable numbers of the SyXB population in the SNC0.25V$_{\rm esc}$1R$_{\rm acc}$ and SND0.25V$_{\rm esc}$1R$_{\rm acc}$ models. We vary the $\alpha$ value to be 0.3 and 5, and compare the results with those in our reference model (with $\alpha=1$) in Table 3. We can see that variation in the $\alpha$ value does not considerably change both numbers. The main reason is that the binding energy of massive star's envelope is so large \citep{Wang2016RAA} that for a low-mass secondary extremely efficient CE ejections requires $\alpha$ to be as large as $\gtrsim 10$ \citep[e.g.,][]{El-Badry2023a}.

In summary, despite the uncertainties in theoretical modeling of SNe and CE evolution, we reach the similar conclusions on the scarcity of BH SyXBs relative to NS SyXBs with various input parameters. However, we caution that the recent discovery of Gaia BHs in wide orbits has strongly challenged the traditional isolated binary evolution theory, suggesting that there may be alternative pathways that led to the formation of BH binaries.

Finally, the detectability of BH SyXBs depends on the structure of the red giant wind. We adopt the traditional BHL accretion mode for a smooth, spherically symmetric wind, while high-resolution observations reveal complex structures and morphologies such as spirals, arcs, disks, and bipolar outflows in the winds of red giants, which may be also shaped by a companion
star \citep{Mauron2006,Maercker2012,Decin2020,Bollen2022,Danilovich2024}. 
Nevertheless, wind-captured disks around the companions are usually observed in high resolution numerical simulations of wind accretion in symbiotic binaries \citep{Espinosa2013,Val-Borro2017,Mellah2020}. Thus, our discussion based on the radiation of disk-accreting BHs may not unreasonable. In addition, \citet{Mohamed2007,Mohamed2010} proposed wind Roche-lobe overflow (WRLOF) as a possible mass transfer mode in symbiotic binaries to explain the the formation of Ba stars and bipolar planetary nebulae, and mass outflow in Mira binaries. WRLOF occurs when the dust-driven winds of a redgiant star fills its Roche lobe and flows onto the companion through the inner Lagrangian point. \citet{2024ApJ...969....8S} recently used WRLOF to account for the characteristics observed in blue lurkers and blue stragglers. However, it is unclear whether WRLOF occurs in SyXBs, since the accretion rate through WRLOF may exceed $\sim 100$ times the BHL accretion rate \citep{Mohamed2007}, which could cause significantly higher X-ray luminosities of SyXBs than observed. Neverless, if there is WRLOF in SyXBs, accretion disks would be formed around NSs and BHs. This would not significantly influence the detectable number of BH SyXBs (because most of them are already considered to be disk-fed) but enhance the number of detectable NS SyXBs. Then, the number ratio of NS and BH SyXBs would be even higher.

\begin{deluxetable*}{cccc}
\centering
\tablenum{3}
\tablecaption{BH SyXB numbers with different $\alpha$ value}
\tabletypesize{\small}
\tablehead{
 $\alpha$  & Model   &  Total numbers  &  Detectable numbers     }
\startdata
	\multirow{2}{*}{0.3}  & SNC0.25V$_{\rm esc}$1R$_{\rm acc}$       & 121 & $7-11$  \\
	 & SND0.25V$_{\rm esc}$1R$_{\rm acc}$       & 13 & $0.11-0.27$  \\
 \hline
    \multirow{2}{*}{1} & SNC0.25V$_{\rm esc}$1R$_{\rm acc}$    & 151 & $8-18$  \\
    &  SND0.25V$_{\rm esc}$1R$_{\rm acc}$     & 7 & $0.18-0.45$  \\
\hline
    \multirow{2}{*}{5} & SNC0.25V$_{\rm esc}$1R$_{\rm acc}$   & 190 & $9-18$  \\
    &   SND0.25V$_{\rm esc}$1R$_{\rm acc}$     & 10 & $0.14-0.33$ \\
\enddata
\end{deluxetable*}

\begin{acknowledgements}
This work was supported by the National Key Research and Development Program of China (2021YFA0718500), the Natural Science Foundation of China under grant No. 12041301 and 12121003.
\end{acknowledgements}

\bibliographystyle{aasjournal}
\bibliography{ms}{}

\begin{thebibliography}{}
\expandafter\ifx\csname natexlab\endcsname\relax\def\natexlab#1{#1}\fi
\providecommand{\url}[1]{\href{#1}{#1}}
\providecommand{\dodoi}[1]{doi:~\href{http://doi.org/#1}{\nolinkurl{#1}}}
\providecommand{\doeprint}[1]{\href{http://ascl.net/#1}{\nolinkurl{http://ascl.net/#1}}}
\providecommand{\doarXiv}[1]{\href{https://arxiv.org/abs/#1}{\nolinkurl{https://arxiv.org/abs/#1}}}

\bibitem[{{Abt}(1983)}]{Abt1983}
{Abt}, H.~A. 1983, \araa, 21, 343, \dodoi{10.1146/annurev.aa.21.090183.002015}

\bibitem[{{Adams} {et~al.}(2017){Adams}, {Kochanek}, {Gerke}, {Stanek}, \& {Dai}}]{Adams2017}
{Adams}, S.~M., {Kochanek}, C.~S., {Gerke}, J.~R., {Stanek}, K.~Z., \& {Dai}, X. 2017, \mnras, 468, 4968, \dodoi{10.1093/mnras/stx816}

\bibitem[{{Bahramian} \& {Degenaar}(2023)}]{Bahramian2023}
{Bahramian}, A., \& {Degenaar}, N. 2023, in Handbook of X-ray and Gamma-ray Astrophysics, 120, \dodoi{10.1007/978-981-16-4544-0_94-1}

\bibitem[{{Basinger} {et~al.}(2021){Basinger}, {Kochanek}, {Adams}, {Dai}, \& {Stanek}}]{Basinger2021}
{Basinger}, C.~M., {Kochanek}, C.~S., {Adams}, S.~M., {Dai}, X., \& {Stanek}, K.~Z. 2021, \mnras, 508, 1156, \dodoi{10.1093/mnras/stab2620}

\bibitem[{{Blandford} \& {Begelman}(1999)}]{Blandford1999}
{Blandford}, R.~D., \& {Begelman}, M.~C. 1999, \mnras, 303, L1, \dodoi{10.1046/j.1365-8711.1999.02358.x}

\bibitem[{{Boccioli} \& {Roberti}(2024)}]{Boccioli2024}
{Boccioli}, L., \& {Roberti}, L. 2024, Universe, 10, 148, \dodoi{10.3390/universe10030148}

\bibitem[{{Bollen} {et~al.}(2022){Bollen}, {Kamath}, {Van Winckel}, {De Marco}, {Verhamme}, {Kluska}, \& {Wardle}}]{Bollen2022}
{Bollen}, D., {Kamath}, D., {Van Winckel}, H., {et~al.} 2022, \aap, 666, A40, \dodoi{10.1051/0004-6361/202243429}

\bibitem[{{Bondi} \& {Hoyle}(1944)}]{Bondi1944}
{Bondi}, H., \& {Hoyle}, F. 1944, \mnras, 104, 273, \dodoi{10.1093/mnras/104.5.273}

\bibitem[{{Burrows} {et~al.}(2023){Burrows}, {Vartanyan}, \& {Wang}}]{Burrows2023}
{Burrows}, A., {Vartanyan}, D., \& {Wang}, T. 2023, \apj, 957, 68, \dodoi{10.3847/1538-4357/acfc1c}

\bibitem[{{Burrows} {et~al.}(2024){Burrows}, {Wang}, \& {Vartanyan}}]{Burrows2024}
{Burrows}, A., {Wang}, T., \& {Vartanyan}, D. 2024, \apjl, 964, L16, \dodoi{10.3847/2041-8213/ad319e}

\bibitem[{{Cao} \& {Wang}(2014)}]{Cao2014}
{Cao}, X., \& {Wang}, J.-X. 2014, \mnras, 444, L20, \dodoi{10.1093/mnrasl/slu102}

\bibitem[{{Chakrabarti} {et~al.}(2023){Chakrabarti}, {Simon}, {Craig}, {Reggiani}, {Brandt}, {Guhathakurta}, {Dalba}, {Kirby}, {Chang}, {Hey}, {Savino}, {Geha}, \& {Thompson}}]{Chakrabarti2023}
{Chakrabarti}, S., {Simon}, J.~D., {Craig}, P.~A., {et~al.} 2023, \aj, 166, 6, \dodoi{10.3847/1538-3881/accf21}

\bibitem[{{Couch} {et~al.}(2020){Couch}, {Warren}, \& {O'Connor}}]{Couch2020}
{Couch}, S.~M., {Warren}, M.~L., \& {O'Connor}, E.~P. 2020, \apj, 890, 127, \dodoi{10.3847/1538-4357/ab609e}

\bibitem[{{Danilovich} {et~al.}(2024){Danilovich}, {Malfait}, {Van de Sande}, {Montarg{\`e}s}, {Kervella}, {De Ceuster}, {Coenegrachts}, {Millar}, {Richards}, {Decin}, {Gottlieb}, {Pinte}, {De Beck}, {Price}, {Wong}, {Bolte}, {Menten}, {Baudry}, {de Koter}, {Etoka}, {Gobrecht}, {Gray}, {Herpin}, {Jeste}, {Lagadec}, {Maes}, {McDonald}, {Marinho}, {M{\"u}ller}, {Pimpanuwat}, {Plane}, {Sahai}, {Wallstr{\"o}m}, {Yates}, \& {Zijlstra}}]{Danilovich2024}
{Danilovich}, T., {Malfait}, J., {Van de Sande}, M., {et~al.} 2024, Nature Astronomy, 8, 308, \dodoi{10.1038/s41550-023-02154-y}

\bibitem[{{Davidsen} {et~al.}(1977){Davidsen}, {Malina}, \& {Bowyer}}]{Davidsen1977}
{Davidsen}, A., {Malina}, R., \& {Bowyer}, S. 1977, \apj, 211, 866, \dodoi{10.1086/154996}

\bibitem[{{Davies} \& {Beasor}(2020)}]{Davies2020}
{Davies}, B., \& {Beasor}, E.~R. 2020, \mnras, 496, L142, \dodoi{10.1093/mnrasl/slaa102}

\bibitem[{{de Val-Borro} {et~al.}(2017{\natexlab{a}}){de Val-Borro}, {Karovska}, {Sasselov}, \& {Stone}}]{deVal2017}
{de Val-Borro}, M., {Karovska}, M., {Sasselov}, D.~D., \& {Stone}, J.~M. 2017{\natexlab{a}}, \mnras, 468, 3408, \dodoi{10.1093/mnras/stx684}

\bibitem[{{de Val-Borro} {et~al.}(2017{\natexlab{b}}){de Val-Borro}, {Karovska}, {Sasselov}, \& {Stone}}]{Val-Borro2017}
---. 2017{\natexlab{b}}, \mnras, 468, 3408, \dodoi{10.1093/mnras/stx684}

\bibitem[{{Decin} {et~al.}(2020){Decin}, {Montarg{\`e}s}, {Richards}, {Gottlieb}, {Homan}, {McDonald}, {El Mellah}, {Danilovich}, {Wallstr{\"o}m}, {Zijlstra}, {Baudry}, {Bolte}, {Cannon}, {De Beck}, {De Ceuster}, {de Koter}, {De Ridder}, {Etoka}, {Gobrecht}, {Gray}, {Herpin}, {Jeste}, {Lagadec}, {Kervella}, {Khouri}, {Menten}, {Millar}, {M{\"u}ller}, {Plane}, {Sahai}, {Sana}, {Van de Sande}, {Waters}, {Wong}, \& {Yates}}]{Decin2020}
{Decin}, L., {Montarg{\`e}s}, M., {Richards}, A.~M.~S., {et~al.} 2020, Science, 369, 1497, \dodoi{10.1126/science.abb1229}

\bibitem[{{Deng} \& {Li}(2024)}]{Deng2024b}
{Deng}, Z.-L., \& {Li}, X.-D. 2024, \apj, 971, 54, \dodoi{10.3847/1538-4357/ad5fec}

\bibitem[{{Deng} {et~al.}(2024){Deng}, {Li}, {Shao}, \& {Xu}}]{Deng2024a}
{Deng}, Z.-L., {Li}, X.-D., {Shao}, Y., \& {Xu}, K. 2024, \apj, 963, 80, \dodoi{10.3847/1538-4357/ad2357}

\bibitem[{{Di Carlo} {et~al.}(2024){Di Carlo}, {Agrawal}, {Rodriguez}, \& {Breivik}}]{Di2024}
{Di Carlo}, U.~N., {Agrawal}, P., {Rodriguez}, C.~L., \& {Breivik}, K. 2024, \apj, 965, 22, \dodoi{10.3847/1538-4357/ad2f2c}

\bibitem[{{Doherty} {et~al.}(2017){Doherty}, {Gil-Pons}, {Siess}, \& {Lattanzio}}]{Doherty2017}
{Doherty}, C.~L., {Gil-Pons}, P., {Siess}, L., \& {Lattanzio}, J.~C. 2017, \pasa, 34, e056, \dodoi{10.1017/pasa.2017.52}

\bibitem[{{El-Badry} {et~al.}(2023{\natexlab{a}}){El-Badry}, {Rix}, {Quataert}, {Howard}, {Isaacson}, {Fuller}, {Hawkins}, {Breivik}, {Wong}, {Rodriguez}, {Conroy}, {Shahaf}, {Mazeh}, {Arenou}, {Burdge}, {Bashi}, {Faigler}, {Weisz}, {Seeburger}, {Almada Monter}, \& {Wojno}}]{El-Badry2023a}
{El-Badry}, K., {Rix}, H.-W., {Quataert}, E., {et~al.} 2023{\natexlab{a}}, \mnras, 518, 1057, \dodoi{10.1093/mnras/stac3140}

\bibitem[{{El-Badry} {et~al.}(2023{\natexlab{b}}){El-Badry}, {Rix}, {Cendes}, {Rodriguez}, {Conroy}, {Quataert}, {Hawkins}, {Zari}, {Hobson}, {Breivik}, {Rau}, {Berger}, {Shahaf}, {Seeburger}, {Burdge}, {Latham}, {Buchhave}, {Bieryla}, {Bashi}, {Mazeh}, \& {Faigler}}]{El-Badry2023b}
{El-Badry}, K., {Rix}, H.-W., {Cendes}, Y., {et~al.} 2023{\natexlab{b}}, \mnras, 521, 4323, \dodoi{10.1093/mnras/stad799}

\bibitem[{{El Mellah} {et~al.}(2020){El Mellah}, {Bolte}, {Decin}, {Homan}, \& {Keppens}}]{Mellah2020}
{El Mellah}, I., {Bolte}, J., {Decin}, L., {Homan}, W., \& {Keppens}, R. 2020, \aap, 637, A91, \dodoi{10.1051/0004-6361/202037492}

\bibitem[{{Ertl} {et~al.}(2016){Ertl}, {Janka}, {Woosley}, {Sukhbold}, \& {Ugliano}}]{Ertl2016}
{Ertl}, T., {Janka}, H.~T., {Woosley}, S.~E., {Sukhbold}, T., \& {Ugliano}, M. 2016, \apj, 818, 124, \dodoi{10.3847/0004-637X/818/2/124}

\bibitem[{{Faucher-Gigu{\`e}re} \& {Kaspi}(2006)}]{Faucher2006}
{Faucher-Gigu{\`e}re}, C.-A., \& {Kaspi}, V.~M. 2006, \apj, 643, 332, \dodoi{10.1086/501516}

\bibitem[{{Fryer} {et~al.}(2012){Fryer}, {Belczynski}, {Wiktorowicz}, {Dominik}, {Kalogera}, \& {Holz}}]{Fryer2012}
{Fryer}, C.~L., {Belczynski}, K., {Wiktorowicz}, G., {et~al.} 2012, \apj, 749, 91, \dodoi{10.1088/0004-637X/749/1/91}

\bibitem[{{Gaia Collaboration} {et~al.}(2024){Gaia Collaboration}, {Panuzzo}, {Mazeh}, {Arenou}, {Holl}, {Caffau}, {Jorissen}, {Babusiaux}, {Gavras}, {Sahlmann}, {Bastian}, {Wyrzykowski}, {Eyer}, {Leclerc}, {Bauchet}, {Bombrun}, {Mowlavi}, {Seabroke}, {Teyssier}, {Balbinot}, {Helmi}, {Brown}, {Vallenari}, {Prusti}, {de Bruijne}, {Barbier}, {Biermann}, {Creevey}, {Ducourant}, {Evans}, {Guerra}, {Hutton}, {Jordi}, {Klioner}, {Lammers}, {Lindegren}, {Luri}, {Mignard}, {Nicolas}, {Randich}, {Sartoretti}, {Smiljanic}, {Tanga}, {Walton}, {Aerts}, {Bailer-Jones}, {Cropper}, {Drimmel}, {Jansen}, {Katz}, {Lattanzi}, {Soubiran}, {Th{\'e}venin}, {van Leeuwen}, {Andrae}, {Audard}, {Bakker}, {Blomme}, {Casta{\~n}eda}, {De Angeli}, {Fabricius}, {Fouesneau}, {Fr{\'e}mat}, {Galluccio}, {Guerrier}, {Heiter}, {Masana}, {Messineo}, {Nienartowicz}, {Pailler}, {Riclet}, {Roux}, {Sordo}, {Gracia-Abril}, {Portell}, {Altmann}, {Benson}, {Berthier}, {Burgess}, {Busonero}, {Busso}, {Cacciari}, {C{\'a}novas}, {Carrasco}, {Carry},
  {Cellino}, {Cheek}, {Clementini}, {Damerdji}, {Davidson}, {de Teodoro}, {Delchambre}, {Dell'Oro}, {Fraile Garcia}, {Garabato}, {Garc{\'\i}a-Lario}, {Haigron}, {Hambly}, {Harrison}, {Hatzidimitriou}, {Hern{\'a}ndez}, {Hestroffer}, {Hodgkin}, {Jamal}, {Jevardat de Fombelle}, {Jordan}, {Krone-Martins}, {Lanzafame}, {L{\"o}ffler}, {Lorca}, {Marchal}, {Marrese}, {Moitinho}, {Muinonen}, {Nu{\~n}ez Campos}, {Oreshina-Slezak}, {Osborne}, {Pancino}, {Pauwels}, {Recio-Blanco}, {Riello}, {Rimoldini}, {Robin}, {Roegiers}, {Sarro}, {Schultheis}, {Smith}, {Sozzetti}, {Utrilla}, {van Leeuwen}, {Weingrill}, {Abbas}, {{\'A}brah{\'a}m}, {Abreu Aramburu}, {Ahmed}, {Altavilla}, {{\'A}lvarez}, {Anders}, {Anderson}, {Anglada Varela}, {Antoja}, {Baig}, {Baines}, {Baker}, {Balaguer-N{\'u}{\~n}ez}, {Balog}, {Barache}, {Barros}, {Barstow}, {Bartolom{\'e}}, {Bashi}, {Bassilana}, {Baudeau}, {Becciani}, {Bedin}, {Bellas-Velidis}, {Bellazzini}, {Beordo}, {Bernet}, {Bertolotto}, {Bertone}, {Bianchi}, {Binnenfeld}, {Blanco-Cuaresma},
  {Bland-Hawthorn}, {Blazere}, {Boch}, {Bossini}, {Bouquillon}, {Bragaglia}, {Braine}, {Bratsolis}, {Breedt}, {Bressan}, {Brouillet}, {Brugaletta}, {Bucciarelli}, {Butkevich}, {Buzzi}, {Camut}, {Cancelliere}, {Cantat-Gaudin}, {Capilla Guilarte}, {Carballo}, {Carlucci}, {Carnerero}, {Carretero}, {Carton}, {Casamiquela}, {Casey}, {Castellani}, {Castro-Ginard}, {Ceraj}, {Cesare}, {Charlot}, {Chaudet}, {Chemin}, {Chiavassa}, {Chornay}, {Chosson}, {Cooper}, {Cornez}, {Cowell}, {Crosta}, {Crowley}, {Cruz Reyes}, {Dafonte}, {Dal Ponte}, {David}, {de Laverny}, {De Luise}, {De March}, {de Torres}, {del Peloso}, {Delbo}, {Delgado}, {Delisle}, {Demouchy}, {Denis}, {Dharmawardena}, {Di Giacomo}, {Diener}, {Distefano}, {Dolding}, {Dsilva}, {Enke}, {Fabre}, {Fabrizio}, {Faigler}, {Fatovi{\'c}}, {Fedorets}, {Fern{\'a}ndez-Hern{\'a}ndez}, {Fernique}, {Figueras}, {Fouron}, {Fragkoudi}, {Gai}, {Galinier}, {Garcia-Serrano}, {Garc{\'\i}a-Torres}, {Garofalo}, {Gerlach}, {Geyer}, {Giacobbe}, {Gilmore}, {Girona}, {Giuffrida},
  {Gomboc}, {Gomez}, {Gonz{\'a}lez-Santamar{\'\i}a}, {Gosset}, {Granvik}, {Gregori Barrera}, {Guti{\'e}rrez-S{\'a}nchez}, {Haywood}, {Helmer}, {Hidalgo}, {Hilger}, {Hobbs}, {Hottier}, {Huckle}, {Jim{\'e}nez-Arranz}, {Juaristi Campillo}, {Kaczmarek}, {Kervella}, {Khanna}, {Kontizas}, {Kordopatis}, {Korn}, {K{\'o}sp{\'a}l}, {Kostrzewa-Rutkowska}, {Kruszy{\'n}ska}, {Kun}, {Lambert}, {Lanza}, {Lebreton}, {Lebzelter}, {Leccia}, {Lecoutre}, {Liao}, {Liberato}, {Licata}, {Livanou}, {Lobel}, {L{\'o}pez-Miralles}, {Loup}, {Madar{\'a}sz}, {Mahy}, {Mann}, {Manteiga}, {Marcellino}, {Marchant}, {Marconi}, {Mar{\'\i}n Pina}, {Marinoni}, {Marshall}, {Mart{\'\i}n Lozano}, {Martin Polo}, {Mart{\'\i}n-Fleitas}, {Marton}, {Mascarenhas}, {Masip}, {Mastrobuono-Battisti}, {McMillan}, {Meichsner}, {Merc}, {Messina}, {Millar}, {Mints}, {Mohamed}, {Molina}, {Molinaro}, {Moln{\'a}r}, {Mongui{\'o}}, {Montegriffo}, {Monti}, {Mora}, {Morbidelli}, {Morris}, {Mudimadugula}, {Muraveva}, {Musella}, {Nagy}, {Nardetto}, {Navarrete}, {Oh},
  {Ordenovic}, {Orenstein}, {Pagani}, {Pagano}, {Palaversa}, {Palicio}, {Pallas-Quintela}, {Pawlak}, {Penttil{\"a}}, {Pesciullesi}, {Pinamonti}, {Plachy}, {Planquart}, {Plum}, {Poggio}, {Pourbaix}, {Price-Whelan}, {Pulone}, {Rabin}, {Rainer}, {Raiteri}, {Ramos}, {Ramos-Lerate}, {Ratajczak}, {Re Fiorentin}, {Regibo}, {Reyl{\'e}}, {Ripepi}, {Riva}, {Rix}, {Rixon}, {Robert}, {Robichon}, {Robin}, {Romero-G{\'o}mez}, {Rowell}, {Ruz Mieres}, {Rybicki}, {Sadowski}, {Sagrist{\`a} Sell{\'e}s}, {Sanna}, {Santove{\~n}a}, {Sarasso}, {Sarmiento}, {Sarrate Riera}, {Sciacca}, {S{\'e}gransan}, {Semczuk}, {Shahaf}, {Siebert}, {Slezak}, {Smart}, {Snaith}, {Solano}, {Solitro}, {Souami}, {Souchay}, {Spitoni}, {Spoto}, {Squillante}, {Steele}, {Steidelm{\"u}ller}, {Surdej}, {Szabados}, {Taris}, {Taylor}, {Teixeira}, {Tepper-Garcia}, {Thuillot}, {Tolomei}, {Tonello}, {Torra}, {Torralba Elipe}, {Trabucchi}, {Trentin}, {Tsantaki}, {Turon}, {Ulla}, {Unger}, {Valtchanov}, {Vanel}, {Vecchiato}, {Vicente}, {Villar}, {Weiler}, {Zhao},
  {Zorec}, {Zucker}, {{\v{Z}}upi{\'c}}, \& {Zwitter}}]{Gaia2024}
{Gaia Collaboration}, {Panuzzo}, P., {Mazeh}, T., {et~al.} 2024, \aap, 686, L2, \dodoi{10.1051/0004-6361/202449763}

\bibitem[{{Galloway} {et~al.}(2002){Galloway}, {Sokoloski}, \& {Kenyon}}]{Galloway2002}
{Galloway}, D.~K., {Sokoloski}, J.~L., \& {Kenyon}, S.~J. 2002, \apj, 580, 1065, \dodoi{10.1086/343798}

\bibitem[{{Ge} {et~al.}(2015){Ge}, {Webbink}, {Chen}, \& {Han}}]{Ge2015}
{Ge}, H., {Webbink}, R.~F., {Chen}, X., \& {Han}, Z. 2015, \apj, 812, 40, \dodoi{10.1088/0004-637X/812/1/40}

\bibitem[{{Ge} {et~al.}(2020){Ge}, {Webbink}, {Chen}, \& {Han}}]{Ge2020}
---. 2020, \apj, 899, 132, \dodoi{10.3847/1538-4357/aba7b7}

\bibitem[{{Gerke} {et~al.}(2015){Gerke}, {Kochanek}, \& {Stanek}}]{Gerke2015}
{Gerke}, J.~R., {Kochanek}, C.~S., \& {Stanek}, K.~Z. 2015, \mnras, 450, 3289, \dodoi{10.1093/mnras/stv776}

\bibitem[{{Hinkle} {et~al.}(2006){Hinkle}, {Fekel}, {Joyce}, {Wood}, {Smith}, \& {Lebzelter}}]{Hinkle2006}
{Hinkle}, K.~H., {Fekel}, F.~C., {Joyce}, R.~R., {et~al.} 2006, \apj, 641, 479, \dodoi{10.1086/500350}

\bibitem[{{Hobbs} {et~al.}(2005){Hobbs}, {Lorimer}, {Lyne}, \& {Kramer}}]{Hobbs2005}
{Hobbs}, G., {Lorimer}, D.~R., {Lyne}, A.~G., \& {Kramer}, M. 2005, \mnras, 360, 974, \dodoi{10.1111/j.1365-2966.2005.09087.x}

\bibitem[{{Hoyle} \& {Lyttleton}(1939)}]{Hoyle1939}
{Hoyle}, F., \& {Lyttleton}, R.~A. 1939, Proceedings of the Cambridge Philosophical Society, 35, 405, \dodoi{10.1017/S0305004100021150}

\bibitem[{{Huarte-Espinosa} {et~al.}(2013){Huarte-Espinosa}, {Carroll-Nellenback}, {Nordhaus}, {Frank}, \& {Blackman}}]{Espinosa2013}
{Huarte-Espinosa}, M., {Carroll-Nellenback}, J., {Nordhaus}, J., {Frank}, A., \& {Blackman}, E.~G. 2013, \mnras, 433, 295, \dodoi{10.1093/mnras/stt725}

\bibitem[{{Hunt}(1971)}]{Hunt1971}
{Hunt}, R. 1971, \mnras, 154, 141, \dodoi{10.1093/mnras/154.2.141}

\bibitem[{{Hurley} {et~al.}(2002){Hurley}, {Tout}, \& {Pols}}]{Hurley2002}
{Hurley}, J.~R., {Tout}, C.~A., \& {Pols}, O.~R. 2002, \mnras, 329, 897, \dodoi{10.1046/j.1365-8711.2002.05038.x}

\bibitem[{{Ichimaru}(1977)}]{Ichimaru1977}
{Ichimaru}, S. 1977, \apj, 214, 840, \dodoi{10.1086/155314}

\bibitem[{{I{\l}kiewicz} {et~al.}(2017){I{\l}kiewicz}, {Miko{\l}ajewska}, \& {Monard}}]{IMM2017}
{I{\l}kiewicz}, K., {Miko{\l}ajewska}, J., \& {Monard}, B. 2017, \aap, 601, A105, \dodoi{10.1051/0004-6361/201630021}

\bibitem[{{Illarionov} \& {Sunyaev}(1975)}]{Illarionov1975}
{Illarionov}, A.~F., \& {Sunyaev}, R.~A. 1975, \aap, 39, 185

\bibitem[{{Iorio} {et~al.}(2023){Iorio}, {Mapelli}, {Costa}, {Spera}, {Escobar}, {Sgalletta}, {Trani}, {Korb}, {Santoliquido}, {Dall'Amico}, {Gaspari}, \& {Bressan}}]{Iorio2023}
{Iorio}, G., {Mapelli}, M., {Costa}, G., {et~al.} 2023, \mnras, 524, 426, \dodoi{10.1093/mnras/stad1630}

\bibitem[{{Ivanova} {et~al.}(2013){Ivanova}, {Justham}, {Chen}, {De Marco}, {Fryer}, {Gaburov}, {Ge}, {Glebbeek}, {Han}, {Li}, {Lu}, {Marsh}, {Podsiadlowski}, {Potter}, {Soker}, {Taam}, {Tauris}, {van den Heuvel}, \& {Webbink}}]{Ivanova2013}
{Ivanova}, N., {Justham}, S., {Chen}, X., {et~al.} 2013, \aapr, 21, 59, \dodoi{10.1007/s00159-013-0059-2}

\bibitem[{{Jerkstrand} {et~al.}(2015){Jerkstrand}, {Smartt}, {Sollerman}, {Inserra}, {Fraser}, {Spyromilio}, {Fransson}, {Chen}, {Barbarino}, {Dall'Ora}, {Botticella}, {Della Valle}, {Gal-Yam}, {Valenti}, {Maguire}, {Mazzali}, \& {Tomasella}}]{Jerkstrand2015J}
{Jerkstrand}, A., {Smartt}, S.~J., {Sollerman}, J., {et~al.} 2015, \mnras, 448, 2482, \dodoi{10.1093/mnras/stv087}

\bibitem[{{Kalogera}(1999)}]{Kalogera1999}
{Kalogera}, V. 1999, \apj, 521, 723, \dodoi{10.1086/307562}

\bibitem[{{Kalogera} \& {Webbink}(1998)}]{Kalogera1998}
{Kalogera}, V., \& {Webbink}, R.~F. 1998, \apj, 493, 351, \dodoi{10.1086/305085}

\bibitem[{{Kiel} {et~al.}(2008){Kiel}, {Hurley}, {Bailes}, \& {Murray}}]{Kiel2008}
{Kiel}, P.~D., {Hurley}, J.~R., {Bailes}, M., \& {Murray}, J.~R. 2008, \mnras, 388, 393, \dodoi{10.1111/j.1365-2966.2008.13402.x}

\bibitem[{{Kobulnicky} \& {Fryer}(2007)}]{Kobulnicky2007}
{Kobulnicky}, H.~A., \& {Fryer}, C.~L. 2007, \apj, 670, 747, \dodoi{10.1086/522073}

\bibitem[{{Kochanek} {et~al.}(2008){Kochanek}, {Beacom}, {Kistler}, {Prieto}, {Stanek}, {Thompson}, \& {Y{\"u}ksel}}]{Kochanek2008}
{Kochanek}, C.~S., {Beacom}, J.~F., {Kistler}, M.~D., {et~al.} 2008, \apj, 684, 1336, \dodoi{10.1086/590053}

\bibitem[{{Kochanek} {et~al.}(2024){Kochanek}, {Neustadt}, \& {Stanek}}]{Kochanek2024}
{Kochanek}, C.~S., {Neustadt}, J. M.~M., \& {Stanek}, K.~Z. 2024, \apj, 962, 145, \dodoi{10.3847/1538-4357/ad18d7}

\bibitem[{{Kroupa} {et~al.}(1993){Kroupa}, {Tout}, \& {Gilmore}}]{Kroupa1993}
{Kroupa}, P., {Tout}, C.~A., \& {Gilmore}, G. 1993, \mnras, 262, 545, \dodoi{10.1093/mnras/262.3.545}

\bibitem[{{Kuranov} \& {Postnov}(2015)}]{Kuranov2015}
{Kuranov}, A.~G., \& {Postnov}, K.~A. 2015, Astronomy Letters, 41, 114, \dodoi{10.1134/S1063773715040064}

\bibitem[{{Liu} {et~al.}(2017){Liu}, {Stancliffe}, {Abate}, \& {Matrozis}}]{Liu2017}
{Liu}, Z.-W., {Stancliffe}, R.~J., {Abate}, C., \& {Matrozis}, E. 2017, \apj, 846, 117, \dodoi{10.3847/1538-4357/aa8622}

\bibitem[{{Luna} {et~al.}(2013){Luna}, {Sokoloski}, {Mukai}, \& {Nelson}}]{Luna2013}
{Luna}, G.~J.~M., {Sokoloski}, J.~L., {Mukai}, K., \& {Nelson}, T. 2013, \aap, 559, A6, \dodoi{10.1051/0004-6361/201220792}

\bibitem[{Lü {et~al.}(2012)Lü, Zhu, Postnov, Yungelson, Kuranov, \& Wang}]{Lv2012}
Lü, G.-L., Zhu, C.-H., Postnov, K.~A., {et~al.} 2012, Monthly Notices of the Royal Astronomical Society, 424, 2265, \dodoi{10.1111/j.1365-2966.2012.21395.x}

\bibitem[{{Maercker} {et~al.}(2012){Maercker}, {Mohamed}, {Vlemmings}, {Ramstedt}, {Groenewegen}, {Humphreys}, {Kerschbaum}, {Lindqvist}, {Olofsson}, {Paladini}, {Wittkowski}, {de Gregorio-Monsalvo}, \& {Nyman}}]{Maercker2012}
{Maercker}, M., {Mohamed}, S., {Vlemmings}, W.~H.~T., {et~al.} 2012, \nat, 490, 232, \dodoi{10.1038/nature11511}

\bibitem[{{Mandel} \& {M{\"u}ller}(2020)}]{Mandel2020}
{Mandel}, I., \& {M{\"u}ller}, B. 2020, \mnras, 499, 3214, \dodoi{10.1093/mnras/staa3043}

\bibitem[{{Masetti} {et~al.}(2011){Masetti}, {Munari}, {Henden}, {Page}, {Osborne}, \& {Starrfield}}]{Masetti2011}
{Masetti}, N., {Munari}, U., {Henden}, A.~A., {et~al.} 2011, \aap, 534, A89, \dodoi{10.1051/0004-6361/201117260}

\bibitem[{{Masetti} {et~al.}(2002){Masetti}, {Dal Fiume}, {Cusumano}, {Amati}, {Bartolini}, {Del Sordo}, {Frontera}, {Guarnieri}, {Orlandini}, {Palazzi}, {Parmar}, {Piccioni}, \& {Santangelo}}]{Masetti2002}
{Masetti}, N., {Dal Fiume}, D., {Cusumano}, G., {et~al.} 2002, \aap, 382, 104, \dodoi{10.1051/0004-6361:20011543}

\bibitem[{{Mauron} \& {Huggins}(2006)}]{Mauron2006}
{Mauron}, N., \& {Huggins}, P.~J. 2006, \aap, 452, 257, \dodoi{10.1051/0004-6361:20054739}

\bibitem[{{Mohamed} \& {Podsiadlowski}(2007)}]{Mohamed2007}
{Mohamed}, S., \& {Podsiadlowski}, P. 2007, in Astronomical Society of the Pacific Conference Series, Vol. 372, 15th European Workshop on White Dwarfs, ed. R.~{Napiwotzki} \& M.~R. {Burleigh}, 397

\bibitem[{{Mohamed} \& {Podsiadlowski}(2010)}]{Mohamed2010}
{Mohamed}, S., \& {Podsiadlowski}, P. 2010, in American Institute of Physics Conference Series, Vol. 1314, International Conference on Binaries: in celebration of Ron Webbink's 65th Birthday, ed. V.~{Kalogera} \& M.~{van der Sluys} (AIP), 51--52, \dodoi{10.1063/1.3536409}

\bibitem[{{M{\"u}ller} {et~al.}(2016){M{\"u}ller}, {Viallet}, {Heger}, \& {Janka}}]{Muller2016}
{M{\"u}ller}, B., {Viallet}, M., {Heger}, A., \& {Janka}, H.-T. 2016, \apj, 833, 124, \dodoi{10.3847/1538-4357/833/1/124}

\bibitem[{{Narayan} \& {Yi}(1994)}]{Narayan1994}
{Narayan}, R., \& {Yi}, I. 1994, \apjl, 428, L13, \dodoi{10.1086/187381}

\bibitem[{{Narayan} \& {Yi}(1995)}]{Narayan1995}
---. 1995, \apj, 452, 710, \dodoi{10.1086/176343}

\bibitem[{{Nespoli} {et~al.}(2010){Nespoli}, {Fabregat}, \& {Mennickent}}]{Nespoli2010}
{Nespoli}, E., {Fabregat}, J., \& {Mennickent}, R.~E. 2010, \aap, 516, A94, \dodoi{10.1051/0004-6361/200913410}

\bibitem[{{Neustadt} {et~al.}(2021){Neustadt}, {Kochanek}, {Stanek}, {Basinger}, {Jayasinghe}, {Garling}, {Adams}, \& {Gerke}}]{Neustadt2021}
{Neustadt}, J.~M.~M., {Kochanek}, C.~S., {Stanek}, K.~Z., {et~al.} 2021, \mnras, 508, 516, \dodoi{10.1093/mnras/stab2605}

\bibitem[{{Nieuwenhuijzen} \& {de Jager}(1990)}]{Nieuwenhuijzen1990}
{Nieuwenhuijzen}, H., \& {de Jager}, C. 1990, \aap, 231, 134

\bibitem[{{Nomoto}(1984)}]{Nomoto1984}
{Nomoto}, K. 1984, \apj, 277, 791, \dodoi{10.1086/161749}

\bibitem[{{O'Connor} \& {Ott}(2011)}]{O'Connor2011}
{O'Connor}, E., \& {Ott}, C.~D. 2011, \apj, 730, 70, \dodoi{10.1088/0004-637X/730/2/70}

\bibitem[{{Os{\l}owski} {et~al.}(2011){Os{\l}owski}, {Bulik}, {Gondek-Rosi{\'n}ska}, \& {Belczy{\'n}ski}}]{Oslowski2011}
{Os{\l}owski}, S., {Bulik}, T., {Gondek-Rosi{\'n}ska}, D., \& {Belczy{\'n}ski}, K. 2011, \mnras, 413, 461, \dodoi{10.1111/j.1365-2966.2010.18147.x}

\bibitem[{{Pavlovskii} {et~al.}(2017){Pavlovskii}, {Ivanova}, {Belczynski}, \& {Van}}]{Pavlovskii2017}
{Pavlovskii}, K., {Ivanova}, N., {Belczynski}, K., \& {Van}, K.~X. 2017, \mnras, 465, 2092, \dodoi{10.1093/mnras/stw2786}

\bibitem[{{Paxton} {et~al.}(2011){Paxton}, {Bildsten}, {Dotter}, {Herwig}, {Lesaffre}, \& {Timmes}}]{Paxton2011}
{Paxton}, B., {Bildsten}, L., {Dotter}, A., {et~al.} 2011, \apjs, 192, 3, \dodoi{10.1088/0067-0049/192/1/3}

\bibitem[{{Paxton} {et~al.}(2013){Paxton}, {Cantiello}, {Arras}, {Bildsten}, {Brown}, {Dotter}, {Mankovich}, {Montgomery}, {Stello}, {Timmes}, \& {Townsend}}]{Paxton2013}
{Paxton}, B., {Cantiello}, M., {Arras}, P., {et~al.} 2013, \apjs, 208, 4, \dodoi{10.1088/0067-0049/208/1/4}

\bibitem[{{Paxton} {et~al.}(2015){Paxton}, {Marchant}, {Schwab}, {Bauer}, {Bildsten}, {Cantiello}, {Dessart}, {Farmer}, {Hu}, {Langer}, {Townsend}, {Townsley}, \& {Timmes}}]{Paxton2015}
{Paxton}, B., {Marchant}, P., {Schwab}, J., {et~al.} 2015, \apjs, 220, 15, \dodoi{10.1088/0067-0049/220/1/15}

\bibitem[{{Pejcha} \& {Thompson}(2015)}]{Pejcha2015}
{Pejcha}, O., \& {Thompson}, T.~A. 2015, \apj, 801, 90, \dodoi{10.1088/0004-637X/801/2/90}

\bibitem[{{Podsiadlowski} {et~al.}(2004){Podsiadlowski}, {Langer}, {Poelarends}, {Rappaport}, {Heger}, \& {Pfahl}}]{Podsiadlowski2004}
{Podsiadlowski}, P., {Langer}, N., {Poelarends}, A.~J.~T., {et~al.} 2004, \apj, 612, 1044, \dodoi{10.1086/421713}

\bibitem[{{Podsiadlowski} {et~al.}(2003){Podsiadlowski}, {Rappaport}, \& {Han}}]{Podsiadlowski2003}
{Podsiadlowski}, P., {Rappaport}, S., \& {Han}, Z. 2003, \mnras, 341, 385, \dodoi{10.1046/j.1365-8711.2003.06464.x}

\bibitem[{{Poelarends} {et~al.}(2017){Poelarends}, {Wurtz}, {Tarka}, {Cole Adams}, \& {Hills}}]{Poelarends2017}
{Poelarends}, A. J.~T., {Wurtz}, S., {Tarka}, J., {Cole Adams}, L., \& {Hills}, S.~T. 2017, \apj, 850, 197, \dodoi{10.3847/1538-4357/aa988a}

\bibitem[{{Portegies Zwart} {et~al.}(1997){Portegies Zwart}, {Verbunt}, \& {Ergma}}]{Portegies1997}
{Portegies Zwart}, S.~F., {Verbunt}, F., \& {Ergma}, E. 1997, \aap, 321, 207, \dodoi{10.48550/arXiv.astro-ph/9701037}

\bibitem[{{Reynolds} {et~al.}(2015){Reynolds}, {Fraser}, \& {Gilmore}}]{Reynolds2015}
{Reynolds}, T.~M., {Fraser}, M., \& {Gilmore}, G. 2015, \mnras, 453, 2885, \dodoi{10.1093/mnras/stv1809}

\bibitem[{{Sazonov} {et~al.}(2020){Sazonov}, {Paizis}, {Bazzano}, {Chelovekov}, {Khabibullin}, {Postnov}, {Mereminskiy}, {Fiocchi}, {B{\'e}langer}, {Bird}, {Bozzo}, {Chenevez}, {Santo}, {Falanga}, {Farinelli}, {Ferrigno}, {Grebenev}, {Krivonos}, {Kuulkers}, {Lund}, {Sanchez-Fernandez}, {Tarana}, {Ubertini}, \& {Wilms}}]{Sazonov2020}
{Sazonov}, S., {Paizis}, A., {Bazzano}, A., {et~al.} 2020, \nar, 88, 101536, \dodoi{10.1016/j.newar.2020.101536}

\bibitem[{{Sen} {et~al.}(2024){Sen}, {El Mellah}, {Langer}, {Xu}, {Quast}, \& {Pauli}}]{Sen2024}
{Sen}, K., {El Mellah}, I., {Langer}, N., {et~al.} 2024, arXiv e-prints, arXiv:2406.08596, \dodoi{10.48550/arXiv.2406.08596}

\bibitem[{{Sen} {et~al.}(2021){Sen}, {Xu}, {Langer}, {El Mellah}, {Sch{\"u}rmann}, \& {Quast}}]{Sen2021}
{Sen}, K., {Xu}, X.~T., {Langer}, N., {et~al.} 2021, \aap, 652, A138, \dodoi{10.1051/0004-6361/202141214}

\bibitem[{{Shakura} {et~al.}(2018){Shakura}, {Postnov}, {Kochetkova}, \& {Hjalmarsdotter}}]{Shakura2018}
{Shakura}, N., {Postnov}, K., {Kochetkova}, A., \& {Hjalmarsdotter}, L. 2018, in Astrophysics and Space Science Library, Vol. 454, Astrophysics and Space Science Library, ed. N.~{Shakura}, 331, \dodoi{10.1007/978-3-319-93009-1_7}

\bibitem[{{Shao} \& {Li}(2014)}]{Shao2014}
{Shao}, Y., \& {Li}, X.-D. 2014, \apj, 796, 37, \dodoi{10.1088/0004-637X/796/1/37}

\bibitem[{{Shao} \& {Li}(2018)}]{Shao2018a}
---. 2018, \apj, 867, 124, \dodoi{10.3847/1538-4357/aae648}

\bibitem[{{Shao} \& {Li}(2019)}]{Shao2019}
---. 2019, \apj, 885, 151, \dodoi{10.3847/1538-4357/ab4816}

\bibitem[{{Shao} \& {Li}(2021)}]{Shao2021}
---. 2021, \apj, 920, 81, \dodoi{10.3847/1538-4357/ac173e}

\bibitem[{{Shapiro} \& {Lightman}(1976)}]{Shapiro1976}
{Shapiro}, S.~L., \& {Lightman}, A.~P. 1976, \apj, 204, 555, \dodoi{10.1086/154203}

\bibitem[{{Smartt} {et~al.}(2009){Smartt}, {Eldridge}, {Crockett}, \& {Maund}}]{Smartt2009}
{Smartt}, S.~J., {Eldridge}, J.~J., {Crockett}, R.~M., \& {Maund}, J.~R. 2009, \mnras, 395, 1409, \dodoi{10.1111/j.1365-2966.2009.14506.x}

\bibitem[{{Stone} {et~al.}(1999){Stone}, {Pringle}, \& {Begelman}}]{Stone1999}
{Stone}, J.~M., {Pringle}, J.~E., \& {Begelman}, M.~C. 1999, \mnras, 310, 1002, \dodoi{10.1046/j.1365-8711.1999.03024.x}

\bibitem[{{Sukhbold} {et~al.}(2016){Sukhbold}, {Ertl}, {Woosley}, {Brown}, \& {Janka}}]{Sukhbold2016}
{Sukhbold}, T., {Ertl}, T., {Woosley}, S.~E., {Brown}, J.~M., \& {Janka}, H.~T. 2016, \apj, 821, 38, \dodoi{10.3847/0004-637X/821/1/38}

\bibitem[{{Sun} {et~al.}(2024){Sun}, {Levina}, {Gossage}, {Kalogera}, {Leiner}, {Geller}, \& {Doctor}}]{2024ApJ...969....8S}
{Sun}, M., {Levina}, S., {Gossage}, S., {et~al.} 2024, \apj, 969, 8, \dodoi{10.3847/1538-4357/ad47c1}

\bibitem[{{Tanikawa} {et~al.}(2023){Tanikawa}, {Hattori}, {Kawanaka}, {Kinugawa}, {Shikauchi}, \& {Tsuna}}]{Tanikawa2023}
{Tanikawa}, A., {Hattori}, K., {Kawanaka}, N., {et~al.} 2023, \apj, 946, 79, \dodoi{10.3847/1538-4357/acbf36}

\bibitem[{{van den Heuvel}(2004)}]{Heuvel2004}
{van den Heuvel}, E.~P.~J. 2004, in ESA Special Publication, Vol. 552, 5th INTEGRAL Workshop on the INTEGRAL Universe, ed. V.~{Schoenfelder}, G.~{Lichti}, \& C.~{Winkler}, 185, \dodoi{10.48550/arXiv.astro-ph/0407451}

\bibitem[{{Verbunt} {et~al.}(2017){Verbunt}, {Igoshev}, \& {Cator}}]{Verbunt2017}
{Verbunt}, F., {Igoshev}, A., \& {Cator}, E. 2017, \aap, 608, A57, \dodoi{10.1051/0004-6361/201731518}

\bibitem[{{Vink}(2017)}]{Vink2017}
{Vink}, J.~S. 2017, \aap, 607, L8, \dodoi{10.1051/0004-6361/201731902}

\bibitem[{{Vink} {et~al.}(2001){Vink}, {de Koter}, \& {Lamers}}]{Vink2001}
{Vink}, J.~S., {de Koter}, A., \& {Lamers}, H.~J.~G.~L.~M. 2001, \aap, 369, 574, \dodoi{10.1051/0004-6361:20010127}

\bibitem[{{Wang} {et~al.}(2016{\natexlab{a}}){Wang}, {Jia}, \& {Li}}]{Wang2016MN}
{Wang}, C., {Jia}, K., \& {Li}, X.-D. 2016{\natexlab{a}}, \mnras, 457, 1015, \dodoi{10.1093/mnras/stw101}

\bibitem[{{Wang} {et~al.}(2016{\natexlab{b}}){Wang}, {Jia}, \& {Li}}]{Wang2016RAA}
---. 2016{\natexlab{b}}, Research in Astronomy and Astrophysics, 16, 126, \dodoi{10.1088/1674-4527/16/8/126}

\bibitem[{{Wang} {et~al.}(2024){Wang}, {Zhao}, {Feng}, {Ge}, {Shao}, {Cui}, {Gao}, {Zhang}, {Wang}, {Li}, {Bai}, {Yuan}, {Huang}, {Yuan}, {Zhang}, {Yi}, {Xiang}, {Li}, {Li}, {Zhang}, {Zhang}, {Han}, {Fan}, {Li}, {Chen}, {Liu}, {Meng}, {Liu}, {Zhang}, {Gu}, \& {Liu}}]{Wang2024}
{Wang}, S., {Zhao}, X., {Feng}, F., {et~al.} 2024, Nature Astronomy, \dodoi{10.1038/s41550-024-02359-9}

\bibitem[{{Webbink}(1984)}]{Webbink1984}
{Webbink}, R.~F. 1984, \apj, 277, 355, \dodoi{10.1086/161701}

\bibitem[{{Woosley} \& {Heger}(2015)}]{Woosley2015}
{Woosley}, S.~E., \& {Heger}, A. 2015, \apj, 810, 34, \dodoi{10.1088/0004-637X/810/1/34}

\bibitem[{{Xie} \& {Yuan}(2012)}]{Xie2012}
{Xie}, F.-G., \& {Yuan}, F. 2012, \mnras, 427, 1580, \dodoi{10.1111/j.1365-2966.2012.22030.x}

\bibitem[{{Xu} \& {Li}(2010{\natexlab{a}})}]{Xu2010a}
{Xu}, X.-J., \& {Li}, X.-D. 2010{\natexlab{a}}, \apj, 716, 114, \dodoi{10.1088/0004-637X/716/1/114}

\bibitem[{{Xu} \& {Li}(2010{\natexlab{b}})}]{Xu2010b}
---. 2010{\natexlab{b}}, \apj, 722, 1985, \dodoi{10.1088/0004-637X/722/2/1985}

\bibitem[{{Yuan} \& {Bu}(2010)}]{Yuan2010}
{Yuan}, F., \& {Bu}, D.-F. 2010, \mnras, 408, 1051, \dodoi{10.1111/j.1365-2966.2010.17175.x}

\bibitem[{{Yuan} \& {Narayan}(2004)}]{Yuan2004}
{Yuan}, F., \& {Narayan}, R. 2004, \apj, 612, 724, \dodoi{10.1086/422802}

\bibitem[{{Yuan} \& {Narayan}(2014)}]{Yuan2014}
---. 2014, \araa, 52, 529, \dodoi{10.1146/annurev-astro-082812-141003}

\bibitem[{{Yungelson} {et~al.}(2019){Yungelson}, {Kuranov}, \& {Postnov}}]{Yungelson2019}
{Yungelson}, L.~R., {Kuranov}, A.~G., \& {Postnov}, K.~A. 2019, \mnras, 485, 851, \dodoi{10.1093/mnras/stz467}

\end{thebibliography}

\end{document}